\documentclass[useAMS,usenatbib]{mn2e}
\usepackage{graphicx}
\usepackage{natbib}
\usepackage{enumerate}
\usepackage{amsmath}
\newcommand\lsim{~\lower.5ex\hbox{$\buildrel < \over \sim$}~}
\newcommand\gsim{~\lower.5ex\hbox{$\buildrel > \over \sim$}~}

\title[Excitation of SDWs by convection in discs]
{Excitation of spiral density waves by convection in accretion
discs}
\author[G.~R.~Mamatsashvili and W.~K.~M. Rice] {G.~R.~Mamatsashvili$^{1,2,3}$\thanks{E-mail:
grm@roe.ac.uk} and W.~K.~M. Rice
$^{1}$\\
$^{1}$ SUPA, Institute for Astronomy, University of Edinburgh,
Blackford Hill, Edinburgh EH9 3HJ, Scotland \\
$^{2}$ Abastumani Astrophysical Observatory, Ilia State University,
2a Kazbegi Ave., Tbilisi 0160, Georgia\\
$^3$ Faculty of Exact and Natural Sciences, Tbilisi State
University, 1 Chavchavadze Ave., Tbilisi 0128, Georgia}
\begin{document}

\date{Accepted 2011 June 21. Received 2011 June 21; in
original form 2011 January 31}

\pagerange{\pageref{firstpage}--\pageref{lastpage}} \pubyear{2010}

\maketitle

\label{firstpage}

\begin{abstract}

Motivated by the recent results of \citet{Lesur_Ogilvie10} on the
transport properties of incompressible convection in protoplanetary
discs, in this paper we study the role of compressibility and hence
of another basic mode -- spiral density waves -- in convective
instability in discs. We analyse the linear dynamics of
non-axisymmetric convection and spiral density waves in a Keplerian
disc with superadiabatic vertical stratification using the local
shearing box approach. It is demonstrated that the shear associated
with Keplerian differential rotation introduces a novel phenomenon,
it causes these two perturbation modes to become coupled: during
evolution the convective mode generates (trailing) spiral density
waves and can therefore be regarded as a new source of spiral
density waves in discs. The wave generation process studied here
owes its existence solely to shear of the disc's differential
rotation, and is a special manifestation of a more general linear
mode coupling phenomena universally taking place in flows with an
inhomogeneous velocity profile. We quantify the efficiency of spiral
density wave generation by convection as a function of azimuthal and
vertical wavenumbers of these modes and find that it is maximal and
most powerful when both these length-scales are comparable to the
disc scale height. We also show that unlike the convective mode,
which tends to transport angular momentum inwards in the linear
regime, the spiral density waves transport angular momentum
outwards. Based on these findings, we suggest that in the non-linear
regime spiral density waves generated by convection may play a role
in enhancing the transport of angular momentum due the convective
mode alone, which is actually being changed to outward by
non-linearity, as indicated by above-mentioned recent developments.
\end{abstract}


\section{Introduction}

The main agent responsible for the anomalous outward transport of
angular momentum in protoplanetary accretion discs is widely
recognised to be turbulence due to magnetorotational instability
\citep[MRI,][]{Balbus_Hawley98,Balbus03}. However, in order for the
MRI to operate, the disc must be sufficiently ionised so that the
gas and magnetic field are effectively coupled
\citep{Blaes_Balbus94,
Sano_etal00,Sano_Stone02,Salmeron_Wardle03,Desch04}. In cold and
dense interiors of protoplanetary discs, the necessary level of
ionisation is not typically reached and, as a result, a large
magnetically inactive region --`dead zone' -- develops in the disc
\citep[see e.g.,][]{Gammie96,Stone_etal00,Fromang_etal02}. Several
mechanisms have been put forward to explain (non-magnetic) transport
in the dead zone: penetration/diffusion of turbulent fluctuations
into the dead zone due to the MRI in the surface layers
\citep{Fleming_Stone03,Turner_etal07,Oishi_etal07,Oishi_etal09},
transport due to self-gravity \citep{Armitage_etal01} and due to
various hydrodynamic candidates, such as vortices, spiral density
waves, Rossby wave and baroclinic instabilities
\citep[e.g.,][]{Lovelace_etal99,Li_etal01,
Klahr_Bodenheimer03,Johnson_Gammie05b,Petersen_etal07,
Johnson_Gammie06,Lesur_Papaloizou10}.

Recent developments indicate that yet another mechanism -- thermal
convection in the vertical direction -- can also provide outward
transport of angular momentum. Vertical convection was first
suggested as a source of angular momentum transport in
protoplanetary discs long before the MRI
\citep{Cameron78,Lin_Papaloizou80,
Lin_Papaloizou85,Ruden_Lin86,Ruden_Pollack91}. In these earlier
studies, it was pointed out that the temperature dependence of the
opacity in protoplanetary discs can lead to convective instability
for a wide range of disc temperatures \citep[see also][]{Rafikov07}.
A corresponding effective viscosity was estimated based on a
phenomenological mixing-length prescription and using this,
evolutionary models of discs -- with outward angular momentum
transport driven solely by convective turbulence -- were
constructed. \citet{Ruden_etal88} carried out a linear analysis of
axisymmetric convective modes in discs and, although linear
axisymmetric modes do not produce torques themselves to transport
angular momentum, these authors also estimated -- from the radial
wavelengths and growth rates of the most unstable axisymmetric modes
-- a Shakura-Sunyaev $\alpha \sim 10^{-3}-10^{-2}$, which might be
in the case of non-linear axisymmetric or non-axisymmetric vertical
convection. However, the non-linear development of these
axisymmetric (two-dimensional) convective modes turned out to lead
to inward (i.e., towards the central star) transport of angular
momentum \citep{Kley_etal93,Rudiger_etal02}, which is obviously not
what is required for the accretion of matter from the disc onto the
central star.

To investigate the problem of convective transport more fully,
subsequent studies considered the dynamics of non-axisymmetric
convection in discs with Keplerian differential rotation, or shear.
In this case, shear-induced effects (see below) come into play for
non-axisymmetric perturbations and hence their dynamics is richer
than that of axisymmetric ones. The linear studies of
\citet{Korycansky92} and \citet[][hereafter RG92]{Ryu_Goodman92} in
the shearing box framework showed that growing non-axisymmetric
shearing waves of the convective mode in the trailing phase lead
predominantly to inward transport of angular momentum. However,
\citet{Lin_etal93} argued that this can be changed if the radial
stratification, which is absent in the shearing box, is taken into
account. A stratified background can support radially localised
packets of the convective mode, which are able to provide an outward
flux of angular momentum. Following more general non-linear
three-dimensional simulations of convection by
\citet{Cabot_Pollack92,Cabot96} and \citet{Stone_Balbus96}, however,
did not yield outward angular momentum transport either; the
time-averaged $\alpha$-parameter turned out to be small and
negative, implying inward transport. This led to vertical convection
being regarded as an inefficient mechanism for driving angular
momentum transport and disc's secular evolution. However,
\citet{Klahr_etal99}, using global disc simulations, demonstrated
that convective flow can in fact be non-axisymmetric (i.e., have
azimuthal structure) and pointed out that the small azimuthal extent
of the computational domains and low Reynolds numbers in the
simulations of \citet{Cabot96} and \citet{Stone_Balbus96} had a
tendency to wipe out all azimuthal variations resulting in the near
axisymmetry of convection, which as noted above, is characterised by
inward transport of angular momentum.

Since high-resolution simulations with improved numerical techniques
are becoming increasingly affordable, interest in convective
transport is reviving. \citet{Lesur_Ogilvie10} performed
incompressible simulations of convective instability in the shearing
box and found, contrary to previous results, that at high enough
Rayleigh (Reynolds) numbers ($\gsim 10^6$), the sign of the
corresponding $\alpha$ is reversed to positive, though it still
remains small. This was demonstrated to be a consequence of the
non-axisymmetric structure of convection, which is established at
large Rayleigh numbers. This plays a central role in that it gives
rise to an appreciable pressure-strain correlation tensor that, in
turn, changes the direction of transport from inwards to outwards.
Past simulations mentioned above, being at low Rayleigh (Reynolds)
numbers, had thus been unable to capture the non-axisymmetry of
convective flow. Analogous simulations by \citet{Kapyla_etal10} also
showed that due to non-axisymmetry, convection can transport angular
momentum outwards. However, the specific parameter range considered
by these authors is somewhat different to that of
\cite{Lesur_Ogilvie10}. Although the findings of both these studies
seem encouraging, they should be viewed as tentative and requiring
further corroboration.

All the above-mentioned studies consider perturbation dynamics in a
sheared environment, as the Keplerian differential rotation of
protoplanetary discs is characterised by a strong radial shear of
the azimuthal velocity. It is well known from fluid dynamical
studies that operators governing the linear dynamics of
perturbations in flows with non-uniform kinematics are non-normal
due to the shear of the mean velocity profile, resulting in a number
of linear transient, or finite-time phenomena \citep[see e.g.,][for
an introduction to the
subject]{Trefethen_etal93,Schmid_Henningson01}. The standard modal
approach (i.e., spectral expansion of perturbations in time and
examination of eigenfrequencies), commonly employed in hydrodynamics
\citep{Drazin_Reid81}, describes perturbation behaviour (stability)
only at asymptotically large times; in fact it is unable to capture
the transient dynamics of perturbations at intermediate times. In
other words, predictions of the modal analysis concerning flow
stability are really relevant only to the asymptotic fate of a flow.
Accordingly, new mathematical methods -- broadly known as the
non-modal approach -- had been developed that allow for the
non-normality-induced perturbation dynamics
\citep[e.g.,][]{Schmid_Henningson01}. The non-modal approach leads
to an initial value problem, enabling us to trace a full temporal
evolution of perturbations and in that respect it is advantageous
over the modal approach.

One of the most important transient phenomenon, which is a
consequence of the non-normality, is the coupling and energy
exchange among different perturbation modes with each other and with
the mean flow occurring during a finite time interval in the linear
theory. The linear mode coupling is a general phenomenon intrinsic
to inhomogeneous/shear flows \citep{Chagelishvili_etal97} and plays
an important role in the subsequent non-linear development of
perturbations and largely defines the characteristics of the
resulting non-linear state (turbulence). Because of strong Keplerian
shear, similar non-normality/shear-induced transient processes
should inevitably take place in discs as well. So, for a proper
understanding of energy exchange processes in the non-linear regime
and ultimately of disc turbulence phenomenon, it is necessary to
first analyse in the linear regime all possible couplings and energy
exchange channels among perturbation modes existing in disc flows.
As pointed out above, the modal approach, which has been as widely
employed in studying perturbation dynamics in discs
\citep[e.g.,][]{Narayan_etal87,Lin_etal90,Lubow_Ogilvie98,
Ogilvie_Lubow99,Li_etal03} as in fluid mechanics, does not account
for a finite-time aspect of perturbation dynamics originating from
the shear of disc flow. For this reason, when analysing temporal
evolution of perturbations in discs, a different technique, a
special type of the non-modal approach -- the method of shearing
waves -- is often employed \citep[e.g., RG92;][hereafter
HP09a]{Goldreich_Lynden-Bell65,Bodo_etal05,Johnson_Gammie05a,
Heinemann_Papaloizou09a}, which we also adopt in this paper.

A special manifestation of the shear-induced linear mode coupling
phenomena in discs is the generation of spiral density and
inertia-gravity waves by vortices, which has already been
extensively studied taking into account the presence of other
factors specific to discs: radial and vertical stratifications,
self-gravity, MRI turbulence, etc
\citep{Davis_etal00,Davis02,Tevzadze_etal03,
Bodo_etal05,Johnson_Gammie05a,Johnson_Gammie05b,
Bodo_etal07,Mamatsashvili_Chagelishvili07,Tevzadze_etal08,Mamatsashvili_Rice09,
Heinemann_Papaloizou09a,Heinemann_Papaloizou09b,Tevzadze_etal10}. In
particular, it was shown that an efficient generation of spiral
density waves (SDWs) by vortices occurs when the characteristic
horizontal length-scale of these two perturbation types is of the
order of the disc thickness, which, in turn, implies that
compressibility effects are important at such length-scales. The
role SDWs play in disc dynamics cannot be overestimated. In
particular, it was suggested that gravitational forces due to
stochastic density perturbations associated with SDWs in a turbulent
disc flow, may be important in the migration of low-mass planets
\citep{Nelson_Papaloizou04,Nelson05}. More significantly, SDWs are
able to enhance angular momentum transport rate \citep[see
e.g.,][]{Johnson_Gammie05b,Oishi_etal09}. SDWs also play a role in
the dynamics and angular momentum transport in a dead zone
\citep{Fleming_Stone03,Oishi_etal07,Oishi_etal09}. So, it is crucial
to identify and analyse the generation mechanisms of SDWs.

Our goal in this paper is to investigate yet another manifestation
of shear-induced linear mode coupling phenomena in discs --
generation of SDWs by convection. In a disc with superadiabatic
vertical stratification, together with the convective mode due to a
negative entropy gradient, there also exists a SDW mode due to
compressibility. As we will show, Keplerian shear of the disc's
differential rotation causes the coupling of these two modes in a
way similar to that of coupling of SDWs to vortices mentioned above.
Since normally SDWs contribute to the angular momentum transport
process, this coupling can in turn have important implications for
convective transport as well. So, convection, like vortices, can be
regarded as a new source of SDWs. However, we omit the vortical mode
in this paper, because in the linear theory presented here, an
exponentially growing convective mode is more powerful and dominates
the algebraic growth of the vortical mode (unless the disc is
self-gravitating). Following RG92, we treat a stratified disc in the
local shearing box approximation. We examine in detail how the
generation of the SDW mode by the convective mode occurs during
evolution and quantify its efficiency as a function of azimuthal and
vertical wavelengths of these modes. We demonstrate that the
efficiency of wave excitation is maximal and powerful when both
these wavelengths are comparable to the disc scale height, that is,
for wavelengths at which compressibility is important. Indeed, the
above-mentioned numerical simulations indicate that convective
motions (rolls) actually extend over the entire scale height of the
disc and are characterised by moderate Mach numbers \citep[see
e.g.,][]{Stone_Balbus96}, implying that compressibility and, hence
SDWs, also actively participate in dynamical processes in
convectively unstable discs and their effects must be understood
properly. Finally, we would like to note that SDW generation can
also be seen in the linear analysis of RG92 and also in a related
linear study of the growth of non-axisymmetric shearing waves of the
convective mode by \citet{Brandenburg_Dintrans06}, though these
authors do not specifically focus on and characterise this
phenomenon.

The paper is organised as follows. Physical model and basic
equations are introduced in Section 2. Shear-induced linear coupling
of SDWs and convection is investigated in detail in Section 3.
Angular momentum transport by these two modes is analysed in Section
4. Summary and discussion are presented in Section 5.

\section{Physical Model and Equations}

To study the dynamics of perturbation modes existing in compressible
and stratified gaseous discs with Keplerian rotation, we adopt a
local shearing box approach of \citet{Goldreich_Lynden-Bell65}. In
the shearing box model, disc dynamics is studied in a local
Cartesian reference frame co-rotating with the disc's angular
velocity at some fiducial radius $r_0$ from the central star, so
curvature effects due to cylindrical geometry of the disc are
ignored. In this coordinate frame, the unperturbed differential
rotation of the disc manifests itself as a parallel azimuthal flow
with a linear velocity shear in the radial direction. The Coriolis
force is included to take into account the effects of the coordinate
frame rotation. The vertical component of the gravity force of the
central star is also present, but self-gravity of the disc is
neglected. As a result, we can write down the three-dimensional
shearing box equations of motion
\begin{equation}
\frac{d{\bf u}}{dt}+2\Omega{\bf\hat{z}}\times {\bf
u}+\nabla\left(-q\Omega^2x^2+\frac{1}{2}\Omega^2z^2\right)+\frac{1}{\rho}\nabla
p=0
\end{equation}
and continuity
\begin{equation}
\frac{d\rho}{dt}+\rho{\bf \nabla}\cdot{\bf u}=0.
\end{equation}
Here ${\bf u}=(u_x,u_y,u_z)$ is the velocity of gas in the local
frame; $\rho$ and $p$ are, respectively, the gas density and
pressure; $\Omega$ is the angular velocity of the local rotating
reference frame, which is equal to the disc's angular velocity at
$r_0$: $\Omega(r_0)$; $x,y,z$ are, respectively, the radial,
azimuthal and vertical coordinates; ${\bf \hat{z}}$ is the unit
vector along the vertical direction; $d/dt=\partial/\partial t+({\bf
u}\cdot \nabla)$ is the total time derivative operator. The shear
parameter $q=1.5$ for the Keplerian differential rotation considered
in this paper.

In the absence of heat exchange, the specific entropy $s=c_v {\rm
ln~} (p/\rho^{\gamma})$ of fluid elements is conserved along
streamlines
\begin{equation}
\frac{ds}{dt}=0,
\end{equation}
where the adiabatic index $\gamma=c_p/c_v$ is the ratio of specific
heats $c_p$ and $c_v$ at constant pressure and constant volume,
respectively (for convenience, below we replace $s/c_v \rightarrow
s$). It follows from equations (1-3) that Ertel's theorem of
potential vorticity conservation holds
\begin{equation}
\frac{d}{dt}\frac{(\nabla\times {\bf u}+2\Omega{\bf\hat{z}})\cdot
\nabla s}{\rho}=0,
\end{equation}
which we will make use of later in classifying perturbation modes in
the disc.

\subsection{The equilibrium disc model}

Equations (1-3) have an equilibrium solution that is stationary and
axisymmetric. In this unperturbed state, the velocity field of
Keplerian rotation represents, as noted above, a parallel azimuthal
flow, ${\bf u_0}$, with a constant radial shear $-q\Omega$:
\[
u_{x0}=u_{z0}=0,~~~u_{y0}=-q\Omega x.
\]
In the shearing box model, equilibrium density, $\rho_0$, and
pressure, $p_0$, depend only on the vertical $z-$coordinate and
satisfy the hydrostatic relation
\begin{equation}
g_0\equiv -\frac{1}{\rho_0}\frac{dp_{0}}{dz}=\Omega^2 z.
\end{equation}
Following RG92 and \citet{Brandenburg_Dintrans06}, we assume the
unperturbed disc to be vertically isothermal with a constant
adiabatic sound speed
\begin{equation}
c_s=\sqrt{\gamma\frac{p_0}{\rho_0}}.
\end{equation}
For simplicity, we use the thin disc approximation with constant
gravity \citep[see e.g.,][RG92]{Shu74,Tevzadze_etal03}. In other
words, the vertically varying gravitational acceleration $\Omega^2z$
is replaced with its some height-averaged constant value $g>0$, but
because the former has opposite signs on either side of the disc
midplane $z=0$, we write
\begin{equation}
g_0=\Omega^2z \rightarrow {\rm sign}(z)g.
\end{equation}
Although this simplification is highly idealised, it greatly
facilitates the mathematical treatment of the problem allowing us to
make Fourier analysis of perturbations in the vertical direction too
in addition to horizontal decomposition into shearing waves (see
below). This, in turn, makes it possible to clearly understand the
key effects of shear on the dynamics of perturbation modes with
various vertical and horizontal length-scales in a stratified disc
and avoids the need to solve complex partial differential equations
with respect to time $t$ and the vertical coordinate $z$. In this
case of constant $g$, substituting replacement (7) into equation (5)
and using expression (6) for the vertically constant sound speed, we
find the distribution of the equilibrium density and pressure with
$z$:
\begin{equation}
\frac{\rho_0(z)}{\rho_m}=\frac{p_0(z)}{p_m}= {\rm
exp}\left(-\frac{|z|}{H}\right),
\end{equation}
where
\[
H=\frac{c_s^2}{\gamma g}
\]
is the vertical stratification scale height of the disc, $\rho_m
\equiv \rho_0(0)$ and $p_m \equiv p_0(0)$ are the midplane values of
the equilibrium density and pressure. As mentioned above, $g$ is
some height-averaged value of acceleration and we choose $g^2$ to be
equal to the density-weighted average of $(\Omega^2z)^2$, where as a
weight function we use an exact expression for the equilibrium
density, $\rho_m {\rm exp}(-\gamma \Omega^2z^2/2c_s^2)$, which is
obtained from equations (5) and (6) with a linearly increasing
gravitational acceleration $\Omega^2z$ (RG92). This relates to the
angular velocity through
\[
\Omega^2=\frac{\gamma g^2}{c_s^2}.
\]
The Brunt-V\"{a}is\"{a}l\"{a} frequency squared is defined as
\[
N_0^2\equiv
\frac{g_0}{\rho_0}\left(\frac{1}{c_{s}^2}\frac{dp_0}{dz}-\frac{d\rho_0}{dz}
\right)=\frac{\gamma-1}{\gamma}\Omega^2.
\]
If $\gamma>1$, then $N_0^2>0$ (subadiabatic thermal stratification)
and the equilibrium vertical structure of the disc is convectively
stable. For $\gamma<1$, then $N_0^2<0$ (superadiabatic thermal
stratification) corresponding to a convectively unstable
equilibrium. For $\gamma=1$ (adiabatic thermal stratification),
$N_0^2=0$ and all motions/modes due to buoyancy disappear. Since our
goal here is to investigate the linear coupling between vertical
convection and SDWs and the possible role of this phenomenon in the
disc dynamics and angular momentum transport, we focus, as in RG92,
only on the superadiabatic vertical structure choosing $\gamma=0.8$
throughout this paper. This value of $\gamma$ is close to unity
reflecting the fact that the degree of superadiabaticity is not very
large in protoplanetary discs \citep{Rafikov07}. Strictly speaking,
the case $\gamma<1$ does not make much thermodynamic sense, because
it implies $c_v>c_p$. In spite of this, there is no problem with
equation (3) describing entropy conservation. In principle, the
convectively unstable regime can be modelled without using the
condition $\gamma<1$, if we appropriately choose cooling and heating
functions in the entropy equation (see also RG92). So, a simple
$\gamma$-prescription mimics the basic features of convectively
stable/unstable discs well without introducing additional heating
and cooling functions that might unnecessarily complicate the
situation.

\subsection{Perturbation equations}

Consider now small perturbations to the equilibrium state (8) with
the background flow ${\bf u_0}$. Linearising equations (1-3) about
this state, we obtain the following system governing the
perturbation dynamics
\begin{equation}
\frac{Du_x'}{Dt}=-\frac{1}{\rho_0}\frac{\partial p'}{\partial
x}+2\Omega u_y',
\end{equation}
\begin{equation}
\frac{Du_y'}{Dt}=-\frac{1}{\rho_0}\frac{\partial p'}{\partial
y}+(q-2)\Omega u_x',
\end{equation}
\begin{equation}
\frac{Du_z'}{Dt}=-\frac{1}{\rho_0}\frac{\partial p'}{\partial
z}-g_0\frac{\rho'}{\rho_0},
\end{equation}
\begin{equation}
\frac{D\rho'}{Dt}+\nabla\cdot (\rho_0 {\bf u'})=0,
\end{equation}
\begin{equation}
\frac{Ds'}{Dt}+\frac{\gamma N_0^2}{g_0}u_z'=0,
\end{equation}
where
\[
\frac{D}{Dt} \equiv \frac{\partial}{\partial t}-q\Omega x
\frac{\partial}{\partial y}
\]
and ${\bf u'},~\rho',~p',~s'$ are the perturbed velocity relative to
the background Keplerian shear flow, perturbed density, pressure and
entropy, respectively. From the definition of specific entropy above
we also have the relation
\begin{equation}
s'=\frac{p'}{p_0}-\gamma \frac{\rho'}{\rho_0}.
\end{equation}
The form of equations (9-13) permits a decomposition of the
perturbed quantities into shearing plane waves, or spatial Fourier
harmonics (SFH) with time-dependent amplitudes and phases,
\begin{equation}
\left(\begin{matrix}
u_x'({\bf r},t)\\
u_y'({\bf r},t)\\
u_z'({\bf r},t)\\
s'({\bf r},t) \end{matrix} \right)=\left(\begin{matrix}
\hat{u}_x(t)\\
\hat{u}_y(t)\\
\hat{u}_z(t)\\
\hat{s}(t)
\end{matrix} \right){\rm
exp}\left(\frac{|z|}{2H}\right){\rm exp}({\rm i}k_x(t)x+{\rm
i}k_yy+{\rm i}k_zz),
\end{equation}
\begin{equation}
\left(\begin{matrix}
\rho'({\bf r},t)\\
p'({\bf r},t) \end{matrix} \right)=\left(\begin{matrix}
\hat{\rho}(t)\\
\hat{p}(t) \end{matrix} \right){\rm
exp}\left(-\frac{|z|}{2H}\right){\rm exp}({\rm i}k_x(t)x+{\rm
i}k_yy+{\rm i}k_zz),
\end{equation}
\[
k_x(t)=q\Omega k_y t.
\]
The azimuthal $k_y$ and vertical $k_z$ wavenumbers remain unchanged,
whereas the radial wavenumber $k_x(t)$ varies with time at a
constant rate $q\Omega k_y$ if $k_y\neq 0$ (i.e., for
non-axisymmetric perturbations) due to sweeping of wave crests by
the background shear flow. For convenience, in $k_x(t)$ we have
shifted the origin of time towards negative values, so that for
$t=0$, $k_x(0)=0$ as well. In this case, an initially leading SFH
with $k_x(t)/k_y < 0$ at $t<0$ eventually becomes trailing with
$k_x(t)/k_y > 0$ at $t>0$. This change of SFH's orientation from
leading to trailing is called `swing' and occurs when $k_x(t)=0$.
This technique of decomposition of perturbations into shearing waves
was originally devised by Lord Kelvin \citep{Thompson87} in order to
study transiently growing solutions in inviscid incompressible
parallel flows with linear shear. It is widely used today in various
applications involving flows with a linear shear of velocity profile
and greatly helps to grasp intermediate-time linear phenomena --
transient growth and coupling of perturbation modes -- in shear
flows, which tend to be overlooked in the standard modal analysis.
Besides, such shearing waves are really the only spectral basis
compatible with shearing-periodic boundary conditions, which are
almost always used in local simulations of accretion discs
\citep[see e.g.,][]{Hawley_etal95}. We would also like to mention
that recently it has been mathematically proven by \citet{Yoshida05}
that shearing waves in fact represent the simplest/basic `elements'
of dynamical processes at linear shear. The exponential factors in
(15) and (16) involving $\pm|z|/2H$ are necessary in order to
compensate for otherwise exponentially increasing with height
perturbation energy due to stratified equilibrium. With the latter
form, the perturbation energy, proportional to $\rho_0u'^2$, is
vertically uniform when averaged over the vertical wavelength
\citep[see also][RG92]{Lerche_Parker67}. The inclusion of these
exponential factors also ensure that there are no imaginary terms in
the dispersion relation derived below that could lead us to infer
some sort of spurious instability.

Substituting (15) and (16) into equations (9-14) and using
replacement (7), we arrive at the following system of first order
ordinary differential equations that govern the linear dynamics of
SFHs of perturbations
\begin{equation}
\frac{d\hat{u}_x}{dt}=-{\rm i}k_x(t)\frac{\hat{p}}{\rho_m}+2\Omega
\hat{u}_y,
\end{equation}
\begin{equation}
\frac{d\hat{u}_y}{dt}=-{\rm
i}k_y\frac{\hat{p}}{\rho_m}+(q-2)\Omega\hat{u}_x,
\end{equation}
\begin{equation}
\frac{d\hat{u}_z}{dt}=-\left({\rm i}k_z-\frac{{\rm
sign}(z)}{2H}\right) \frac{\hat{p}}{\rho_m}-{\rm
sign}(z)g\frac{\hat{\rho}}{\rho_m},
\end{equation}
\begin{equation}
\frac{d\hat{\rho}}{dt}=-{\rho_m}\left[{\rm i}k_x(t)\hat{u}_x+{\rm
i}k_y\hat{u}_y+\left({\rm i}k_z-\frac{{\rm
sign}(z)}{2H}\right)\hat{u}_z\right],
\end{equation}
\begin{equation}
\frac{d\hat{s}}{dt}=-{\rm sign}(z)\frac{\gamma N_0^2}{g}\hat{u}_z,
\end{equation}
\begin{equation}
\hat{s}=\frac{\hat{p}}{p_m}-\gamma \frac{\hat{\rho}}{\rho_m}.
\end{equation}
Note that by means of the exponential factors in (15) and (16), we
have rendered the coefficients in these equations $z$-independent,
though they still contain the sign of $z$. Below we will reduce this
set so that sign-dependence will disappear too.

It can be readily shown that the system (17-21) possesses an
important time invariant -- the linearised potential vorticity
perturbation
\begin{multline*}
I\equiv {\rm i}k_x(t)\hat{u}_y-{\rm i}k_y\hat{u}_x-\\
-(2-q)\Omega\left[\frac{\hat{\rho}}{\rho_m}-\frac{{\rm
sign}(z)g}{\gamma N_0^2}\left({\rm i} k_z-\frac{{\rm
sign}(z)}{2H}\right)\hat{s}\right]=const,
\end{multline*}
whose conservation is a direct consequence of Ertel's potential
vorticity theorem (4). This time invariant $I$, in turn, indicates
the existence of the vortical/aperiodic mode in the perturbation
spectrum, which is characterised by non-zero potential vorticity
($I\neq 0$) and represents a stationary (i.e., time-independent)
solution of equations (17-21) in the absence of shear. In the
present three-dimensional case, this vortical mode originates from
the combined action of the vertical gravity force (stratification)
and the Coriolis force and represents a 3D generalisation of the 2D
vortical mode considered in
\citet[][]{Bodo_etal05,Johnson_Gammie05a,Mamatsashvili_Chagelishvili07};
HP09a. Because of the shear/non-normality of disc flow, the vortical
mode can undergo large transient amplification and effectively
(linearly) couple with and excite other modes (SDWs, inertia-gravity
waves, baroclinic modes, etc.) existing in the disc
\citep[][HP09a]{Tevzadze_etal03,
Bodo_etal05,Mamatsashvili_Chagelishvili07,Tevzadze_etal08,
Tevzadze_etal10}. However, in the linear regime considered here with
no disc self-gravity, the transient (algebraic) growth of the
vortical mode is overwhelmed by the exponential amplification of the
convective mode, so we exclude the former mode from our analysis by
setting the linearised potential vorticity to zero, $I=0$. In the
non-linear regime, however, the exponential growth of the convective
mode is saturated, so convection no longer dominates over vortices
and both can equally take part in the SDW generation process.

Taking into account that by our choice $I=0$, we can eliminate
$\hat{u}_x,~\hat{u}_y,~\hat{u}_z$ and $\hat{p}$ in favour of
$\hat{\rho}$ and $\hat{s}$ in equations (17-22). Then, for
convenience, switching to a new auxiliary quantity
\begin{multline*}
\hat{l}=\frac{1}{k_{\perp}(t)}\left[\frac{1}{\gamma}+\frac{{\rm
sign}(z)g}{\gamma N_0^2}\left({\rm i}k_z-\frac{{\rm
sign}(z)}{2H}\right)\right]^{-1}\times \\
\times\left[\frac{\hat{\rho}}{\rho_m}-\frac{{\rm sign}(z)g}{\gamma
N_0^2}\left({\rm i}k_z-\frac{{\rm sign}(z)}{2H}\right)\hat{s}\right]
\end{multline*}
instead of the density $\hat{\rho}$, we finally arrive at the second
order system with real coefficients forming the basis for our
subsequent analysis,
\begin{equation}
\frac{d^2\hat{l}}{dt^2}=-[c_s^2k_{\perp}^2(t)+\kappa^2(t)]\hat{l}-c_s^2k_{\perp}(t)\hat{s},
\end{equation}
\begin{equation}
\frac{d^2\hat{s}}{dt^2}=k_{\perp}(t)\left[N_0^2-c_s^2\left(k_z^2+\frac{1}{4H^2}\right)\right]\hat{l}-c_s^2\left(k_z^2+\frac{1}{4H^2}\right)\hat{s},
\end{equation}
where
\[
\kappa^2(t)\equiv
2(2-q)\Omega^2-\frac{4q\Omega^2k_y^2}{k_{\perp}^2(t)}+\frac{3q^2\Omega^2k_y^4}{k_{\perp}^4(t)}
\]
and $k_{\perp}^2(t)=k_x^2(t)+k_y^2=k_y^2(1+q^2\Omega^2t^2)$.
Although the physical meaning of $\hat{l}$ is less transparent, by
taking its advantage we have reduced the original system (17-22) to
a more compact one (23) and (24), with all the coefficients being
real, in order to ease further manipulations and to make our
equations friendlier for numerical treatment. Equations (23) and
(24) contain all the information on various perturbation modes
(apart from the vortical mode) present in a vertically stratified
Keplerian disc. In the next section, we remove shear from these
equations for the moment in order to classify modes in the disc and
then put shear back again to see how it alters the mode dynamics and
what new effects it introduces.

\subsection{Classification of modes in the absence of shear}

Consider first a simple case without shear (i.e., a rigidly rotating
disc) by setting $q=0$ in equations (23) and (24). After that, all
the coefficients in these equations become time-independent and,
therefore, we can look for solutions in the form
$\hat{l},\hat{s}\propto {\rm exp}({\rm i}\omega t)$. Substituting
this into equations (23) and (24), we obtain the following
dispersion relation in the absence of shear
\begin{multline*}
\omega^4-\left[c_s^2k_{\perp}^2+c_s^2\left(k_z^2+\frac{1}{4H^2}\right)+\kappa^2\right]\omega^2+\\
+c_s^2\left[N_0^2k_{\perp}^2+\kappa^2\left(k_z^2+\frac{1}{4H^2}\right)\right]=0,
\end{multline*}
where $\kappa^2=4\Omega^2$ is constant when $q=0$.
This dispersion relation has two different solutions corresponding
to two different types of perturbation modes:\footnote{If we
included the vortical mode, it would correspond, as noted above, to
the stationary solution $\omega=0$, since $I\neq 0$ for this mode.}
\begin{enumerate}[1.]
\item
A high-frequency SDW mode with
\begin{multline}
\omega_s^2=\frac{c_s^2k_{\perp}^2+c_s^2(k_z^2+1/4H^2)+\kappa^2}{2}\times\\
\times\left(1+\sqrt{1-\frac{4c_s^2[N_0^2k_{\perp}^2+
\kappa^2(k_z^2+1/4H^2)]}{[c_s^2k_{\perp}^2+c_s^2(k_z^2+1/4H^2)+\kappa^2]^2}}\right)
\end{multline}
the restoring force for which is mainly provided by
compressibility/pressure forces, but is modified by
stratification/bouyancy and the Coriolis force. So, in this mode,
pressure perturbations dominate over entropy perturbations. For
large wavenumbers $k_{\perp}H, k_zH \gg 1$ (i.e., for wavelengths
much smaller than the disc scale height), the effects of
stratification and rotation are negligible and the frequency of the
SDW mode reduces to $\omega_s^2\simeq c_s^2(k_{\perp}^2+k_z^2)$. In
an unstratified disc, for $z-$independent perturbations (i.e., in
the razor-thin disc approximation), from (25) we obtain
$\omega_s^2=c_s^2(k_x^2+k_y^2)+\kappa^2$, which is a classical
dispersion relation of two-dimensional SDWs in non-self-gravitating
discs \citep[see e.g.,][]{Balbus03} and therefore expression (25)
can be regarded as a generalised form of the SDW mode frequency in
the presence of vertical stratification.
\item
A low-frequency inertia-buoyancy mode with
\begin{multline}
\omega_s^2=\frac{c_s^2k_{\perp}^2+c_s^2(k_z^2+1/4H^2)+\kappa^2}{2}\times\\
\times\left(1-\sqrt{1-\frac{4c_s^2[N_0^2k_{\perp}^2+
\kappa^2(k_z^2+1/4H^2)]}{[c_s^2k_{\perp}^2+c_s^2(k_z^2+1/4H^2)+\kappa^2]^2}}\right)
\end{multline}
the restoring force for which is mainly provided by vertical
buoyancy and the Coriolis force, but modified by compressibility.
So, in this mode, entropy perturbations dominate over pressure
perturbations. The inertia-buoyancy mode represents an
inertia-gravity wave in the case of subadiabatic and adiabatic
stratifications ($N_0^2\geq 0$) and the convective mode in the case
of superadiabatic stratification ($N_0^2<0$), which is considered
here.\footnote{Although we call this a convective mode, it actually
exhibits convective instability (i.e., it has $\omega_g^2<0$) only
when the wavenumbers satisfy
$\kappa^2(k_z^2+1/4H^2)<-N_0^2k_{\perp}^2$ in order to overcome the
stabilising effect of rotation.} The dynamics of inertia-gravity
waves in Keplerian discs is extensively studied elsewhere \citep[see
e.g.,][]{Lubow_Pringle93,Ogilvie98,Balbus03,
Tevzadze_etal03,Tevzadze_etal08,Latter_Balbus09} and will not be
dealt with here. Again, in the limit of large wavenumbers
$k_{\perp}H, k_zH \gg 1$, the effect of compressibility on the
inertia-buoyancy mode is small and we get the well-known dispersion
relation $\omega_g^2\simeq
\frac{N_0^2k_{\perp}^2+\kappa^2k_z^2}{k_{\perp}^2+k_z^2}$, which can
also be derived using the incompressible (Boussinesq) approximation
\citep{Tevzadze_etal08}.
\end{enumerate}
Thus, in a stratified compressible disc, perturbations can be
classified into two basic types -- the SDW and inertia-buoyancy
modes (apart from the vortical mode) and any general perturbation
(with zero potential vorticity) can be decomposed into the sum of
these two modes. We should note, however, that the mode
classification performed here is actually applicable in the case of
the constant-$g$ approximation (7), which is, strictly speaking,
valid when characteristic vertical length-scales of perturbations
are less than the disc scale height. A more general classification
of vertical modes in stratified discs without invoking this
approximation is performed in \citet{Ruden_etal88,Lubow_Pringle93,
Korycansky_Pringle95,Ogilvie98} (in these papers, the compressible
p-mode corresponds to our SDW mode and the r- and convective g-modes
to the inertia-buoyancy mode). Nevertheless, the classification
adopted here allows us to grasp the key effects of shear on the
dynamics of the inertia-buoyancy (convective) and SDW modes.

To analyse the behaviour of these two modes, we define new
characteristic quantities, or eigenfunctions, $\psi_s$ and $\psi_g$,
for them as
\begin{equation}
\psi_s=\frac{k_{\perp}[c_s^2(k_z^2+1/4H^2)-N_0^2]\hat{l}-[\omega_g^2-c_s^2(k_z^2+1/4H^2)]\hat{s}}{\omega_s^2-\omega_g^2},
\end{equation}
\begin{equation}
\psi_g=\frac{k_{\perp}[c_s^2(k_z^2+1/4H^2)-N_0^2]\hat{l}-[\omega_s^2-c_s^2(k_z^2+1/4H^2)]\hat{s}}{\omega_g^2-\omega_s^2}
\end{equation}
describing, respectively, the SDW and inertia-buoyancy modes. These
eigenfunctions are convenient and physically revealing, as
substituting them into equations (23) and (24) with shear terms
removed, we obtain two separate equations for each mode
eigenfunction
\begin{equation}
\frac{d^2\psi_s}{dt^2}+\omega_s^2\psi_s=0,~~~~
\frac{d^2\psi_g}{dt^2}+\omega_g^2\psi_g=0
\end{equation}
that allow us to study these modes individually. All the other
quantities ($\hat{u}_x, \hat{u}_y, \hat{u}_z, \hat{\rho}, \hat{p}$)
can be expressed through the mode eigenfunctions $\psi_s,\psi_g$ and
their respective first order time-derivatives. Hence, we can fully
determine the perturbation field corresponding to a specific mode by
setting the eigenfunction of the other mode and its time derivative
to zero. In other words, if we want to have either only SDWs or the
inertia-buoyancy mode in the disc flow, we should initially set
simply $\psi_g(-\infty)=d\psi_g/dt(-\infty)=0$ or
$\psi_s(-\infty)=d\psi_s/dt(-\infty)=0$, respectively. Modal
equations (29) governing the dynamics of the SDW and
inertia-buoyancy modes are not coupled, implying that in the absence
of shear the perturbation modes evolve independently, that is,
initially exciting one either mode with a specific characteristic
time-scale does not lead to the excitation of other modes with
different time-scales. In the following sections, we investigate how
shear modifies modal equations (29) and its influence on the
dynamics of the SDW and inertia-buoyancy modes. Specifically, we
will show that in the presence of Keplerian shear, these equations
are no longer independent from each other, i.e., become coupled
that, in turn, results in the linear coupling between these two
modes and, in particular, between SDWs and convection.

\subsection{Effects of shear on mode dynamics}

Here we derive the generalised modal equations from equations (23)
and (24) including shear $q$. For the sake of further analysis, let
us first non-dimensionalise the basic quantities by obvious units:
time (frequencies) by means of the angular speed $\Omega$,
\[
\Omega t \rightarrow t,~~~\frac{\omega_{s,g}}{\Omega}\rightarrow
\omega_{s,g},~~~\frac{\kappa^2}{\Omega^2} \rightarrow
\kappa^2,~~~\frac{N_0^2}{\Omega^2}\rightarrow
N_0^2=\frac{\gamma-1}{\gamma},
\]
and length-scales (wavenumbers) as well as $\hat{l}$ using
$c_s/\Omega$ as a unit of length,
\[
(K_x, K_y, K_z) = \frac{c_s}{\Omega}(k_x, k_y, k_z),~~\frac{\Omega
H}{c_s}\rightarrow H=\frac{1}{\sqrt{\gamma}},~~\frac{\Omega
\hat{l}}{c_s}\rightarrow \hat{l}
\]
($\hat{s}, \psi_s, \psi_g$ are already non-dimensional).
Accordingly, the normalised radial wavenumber $K_x$ varies with
normalised time as $K_x(t)=qK_yt$. Also, as defined above,
$K_{\perp}^2(t)=K_x^2(t)+K_y^2=K_y^2(1+q^2t^2)$. Since $K_x(t)$ and
$\kappa^2(t)$ in equations (25) and (26) are now time-dependent as a
result of shear, the non-dimensional frequencies $\omega_s$ and
$\omega_g$ also become functions of time:
\begin{multline*}
\omega_s^2(t)=\frac{\kappa^2(t)+K_{\perp}^2(t)+K_z^2+\gamma/4}{2}\times\\
\times\left(1+\sqrt{1-\frac{4[N_0^2K_{\perp}^2(t)+
\kappa^2(t)(K_z^2+\gamma/4)]}{[\kappa^2(t)+K_{\perp}^2(t)+K_z^2+\gamma/4]^2}}\right)
\end{multline*}
\begin{multline*}
\omega_g^2(t)=\frac{\kappa^2(t)+K_{\perp}^2(t)+K_z^2+\gamma/4}{2}\times\\
\times\left(1-\sqrt{1-\frac{4[N_0^2K_{\perp}^2(t)+
\kappa^2(t)(K_z^2+\gamma/4)]}{[\kappa^2(t)+K_{\perp}^2(t)+K_z^2+\gamma/4]^2}}\right).
\end{multline*}
At large times, $K_{\perp}(t)\rightarrow \infty$ and from these
expressions it follows that $\omega_s^2(t)\simeq
K_{\perp}^2(t),~\omega_g^2(t)\simeq N_0^2$. As noted above, all the
other quantities can readily be expressed through $\psi_s,\psi_g$
and their first order time derivatives, so we carry out the further
analysis working mostly with the eigenfunctions.

Expressing $\hat{l},\hat{s}$ through $\psi_s,\psi_g$ from equations
(27) and (28) and substituting them into equations (23) and (24), we
arrive at the following system of coupled second order differential
equations for the eigenfunctions
\begin{equation}
\frac{d^2\psi_s}{dt^2}+f_s\frac{d\psi_s}{dt}
+(\omega_s^2+\Delta\omega_s^2)\psi_s=f_g\frac{d\psi_g}{dt}+\Delta\omega_g^2\psi_g,
\end{equation}
\begin{equation}
\frac{d^2\psi_g}{dt^2}+f_g\frac{d\psi_g}{dt}
+(\omega_g^2+\Delta\omega_g^2)\psi_g=f_s\frac{d\psi_s}{dt}+\Delta\omega_s^2\psi_s,
\end{equation}
where the new coefficients $f_s,\Delta\omega_s^2$ and
$f_g,\Delta\omega_g^2$, compared with equations (29), describe
modifications to the dynamics of the SDW and inertia-buoyancy modes
and their coupling brought about by the shear of disc flow. They
depend on time as well as on the azimuthal $K_y$ and vertical $K_z$
wavenumbers:
\[
f_s(K_y,K_z,
t)\equiv\frac{2K_{\perp}(t)}{\omega_s^2(t)-\omega_g^2(t)}\left(\frac{\omega_s^2(t)-K_z^2-\gamma/4}{K_{\perp}(t)}\right)',
\]
\[
f_g(K_y,K_z,
t)\equiv\frac{2K_{\perp}(t)}{\omega_g^2(t)-\omega_s^2(t)}\left(\frac{\omega_g^2(t)-K_z^2-\gamma/4}{K_{\perp}(t)}\right)',
\]
\[
\Delta\omega_s^2(K_y,K_z,
t)\equiv\frac{K_{\perp}(t)}{\omega_s^2(t)-\omega_g^2(t)}\left(\frac{\omega_s^2(t)-K_z^2-\gamma/4}{K_{\perp}(t)}\right)'',
\]
\[
\Delta\omega_g^2(K_y,K_z,
t)\equiv\frac{K_{\perp}(t)}{\omega_g^2(t)-\omega_s^2(t)}\left(\frac{\omega_g^2(t)-K_z^2-\gamma/4}{K_{\perp}(t)}\right)''
\]
(here and below primes denote the time derivative). Note that all
the time derivatives in these coefficients are proportional to the
shear parameter $q$ as well as to the azimuthal wavenumber $K_y$
(time enters through $qK_yt$) and thus vanish in the shearless limit
$q=0$ or in the case of axisymmetric perturbations with $K_y=0$,
reducing equations (30) and (31) to decoupled equations (29). In
vertically stratified Keplerian discs, the dynamics of axisymmetric
perturbations (modes) was extensively studied in the past
\citep{Ruden_etal88,Kley_etal93,Lubow_Pringle93,Korycansky_Pringle95,
Ogilvie98,Mamatsashvili_Rice10}, so we omit them from the present
analysis and concentrate only on non-axisymmetric perturbations
having non-zero azimuthal wavenumber $K_y\neq 0$, which we assume to
be positive ($K_y>0$) throughout without loss of generality.

Equations (30) and (31) describe the linear dynamics of the SDW and
inertia-buoyancy modes and their coupling in a vertically
stratified, compressible disc with (Keplerian) differential
rotation. As mentioned above, novel features in the mode dynamics,
in comparison with the case of no shear, arise from the coefficients
$f_s, \Delta\omega_s^2$ and $f_g, \Delta\omega_g^2$ that originate
from shear. The homogeneous (left-hand side) parts of equations (30)
and (31) describe the individual behaviour of the SDW and
inertia-buoyancy modes, respectively, in the presence of shear. The
most interesting aspect here is the source terms on the right-hand
side of these equations that provide coupling between these two
modes. In other words, initially imposed only one mode acts as a
source for another mode and can excite it in the course of
evolution. Thus, the shear of disc flow introduces a phenomenon of
mode conversion. We will see below that because of such a coupling,
the convective and SDW modes actually have strictly separate
identities only at large times ($|t|\gg 1$), when the characteristic
time-scales (frequencies/growth rates) of these modes differ
significantly (Fig. 1). As seen from Fig. 1, for finite times ($|t|
\lsim 1$) instead the mode time-scales are comparable and, as a
result, the modes cannot be quite disentangled from each other, so
that we have some mixture -- `convective-SDW' -- mode at such times.
If there was not such a difference in the SDW and convective modes'
time-scales, strictly speaking, it also would not make much sense to
talk about homogeneous and particular inhomogeneous solutions of
equations (30) and (31) separately, which themselves constitute a
homogeneous system. Below, we will demonstrate using numerical
analysis how the linear mode coupling occurs in practice.

For the purpose of numerical integration presented below, we change
the eigenfunctions to $\Psi_s = \psi_s/h_s,~\Psi_g = \psi_g/h_g$,
where
\[
h_s={\rm exp}\left(-\frac{1}{2}\int_{0}^tf_sdt'\right),~~~h_g={\rm
exp}\left(-\frac{1}{2}\int_{0}^tf_gdt'\right).
\]
These new eigenfunctions are more convenient, because substituting
them into equations (30) and (31), we arrive at a simpler system
without first derivatives on the left hand side
\begin{equation}
\frac{d^2\Psi_s}{dt^2}+\hat{\omega}_s^2\Psi_s=\chi_{sg1}\frac{d\Psi_g}{dt}+\chi_{sg2}\Psi_g,
\end{equation}
\begin{equation}
\frac{d^2\Psi_g}{dt^2}+\hat{\omega}_g^2\Psi_g=\chi_{gs1}\frac{d\Psi_s}{dt}+\chi_{gs2}\Psi_s.
\end{equation}
Equations (32) and (33) with time-dependent (modified) frequencies
\[
\hat{\omega}_s^2=\omega_s^2+\Delta\omega_s^2-\frac{f'_s}{2}-\frac{f_s^2}{4},
\]
\[
\hat{\omega}_g^2=\omega_g^2+\Delta\omega_g^2-\frac{f'_g}{2}-\frac{f_g^2}{4},
\]
respectively, for the SDW and inertia-buoyancy modes, resemble
equations of two coupled oscillators. These frequencies differ from
corresponding $\omega_s^2$ and $\omega_g^2$ due to the presence of
the shear-induced terms. For large times as well as for $K_y\gg 1$
and/or $K_z\gg 1$, these terms are small and $\hat{\omega}_s^2$ and
$\hat{\omega}_g^2$ go to corresponding values in the shearless
limit:
\begin{equation*}
\hat{\omega}_s^2(t) \simeq K_{\perp}^2(t)+K_z^2,
\end{equation*}
\begin{equation*}
\hat{\omega}_g^2(t) \simeq
\frac{N_0^2K_{\perp}^2(t)+\kappa^2(t)(K_z^2+\gamma/4)}{K_{\perp}^2(t)+K_z^2}.
\end{equation*}
The $\chi$ parameters, describing the coupling between the modes,
are given by
\[
\chi_{sg1}=\frac{f_gh_g}{h_s},~~~~\chi_{sg2}=\frac{h_g}{h_s}\left(\Delta\omega_g^2-\frac{f_g^2}{2}\right),
\]
\[
\chi_{gs1}=\frac{f_sh_s}{h_g},~~~~\chi_{gs2}=\frac{h_s}{h_g}\left(\Delta\omega_s^2-\frac{f_s^2}{2}\right).
\]
\begin{figure}
\centering\includegraphics[width=\columnwidth]{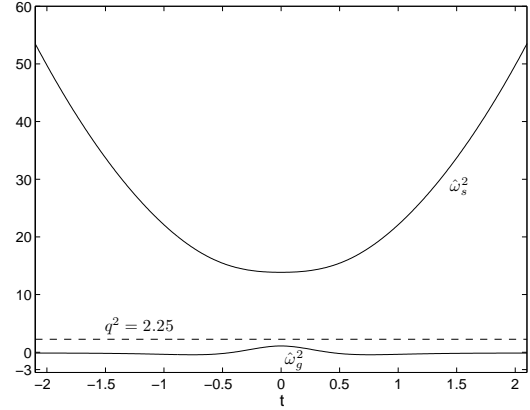}
\caption{Modified frequencies $\hat{\omega}_s^2$ and
$\hat{\omega}_g^2$ as functions of time at $K_y=2,~K_z=3$ for the
convectively unstable stratification with $\gamma=0.8$. For
comparison, with dashed line we also show the square of the inverse
shear (non-dimensional) time $q^2=2.25$. In the vicinity of $t=0$,
all these three time-scales become comparable, whereas at large
times they get separated: $\hat{\omega}_s^2$ increases as $\propto
t^2$ and $\hat{\omega}_g^2$ tends to constant $N_0^2=-0.25$.}
\end{figure}
\begin{figure}
\centering\includegraphics[width=\columnwidth]{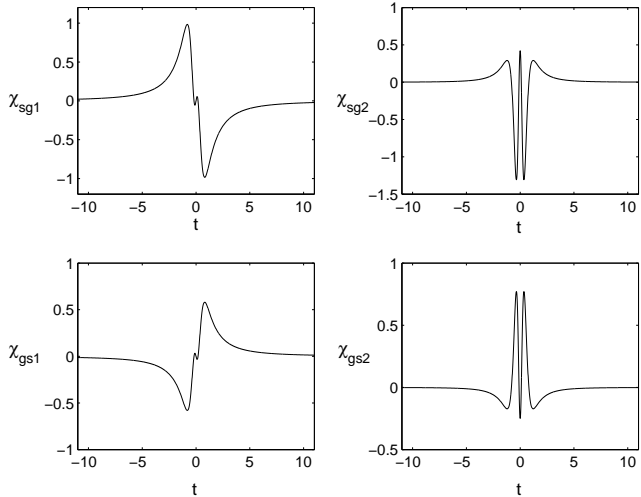}
\caption{Coupling parameters $\chi$ as a function of time for
$K_y=2,~K_z=3$. They reach the highest values during $|t|\lsim 1$
and fall off at large times $|t| \gg 1$.}
\end{figure}
\begin{figure}
\centering\includegraphics[width=\columnwidth]{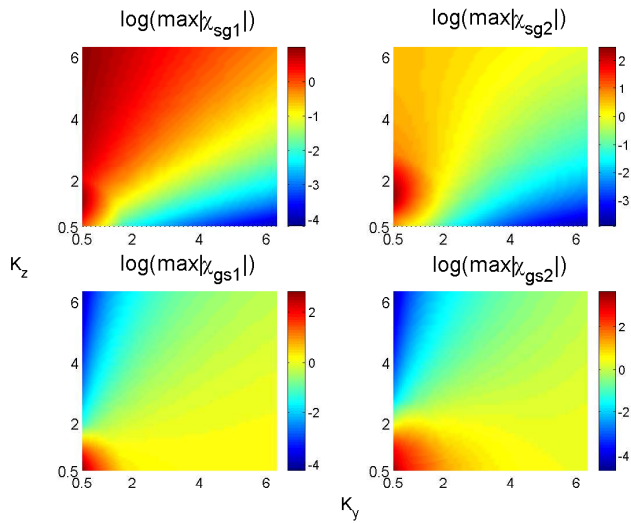}
\caption[Maximal values of modulus of coupling parameters, $|\chi|$,
as a function of $K_y$ and $K_z$]{Maximum values of modulus of
coupling parameters, $|\chi|$, as a function of $K_y$ and $K_z$.
$\max|\chi_{sg1}|$ and $\max|\chi_{sg2}|$ are appreciable over a
broader range of wavenumbers than $\max|\chi_{gs1}|$ and
$\max|\chi_{gs2}|$.}
\end{figure}
The coupling parameters $\chi_{sg1}, \chi_{sg2}$ describe the
excitation of the SDW mode by the inertia-buoyancy mode, while
$\chi_{gs1},\chi_{gs2}$ describe the excitation of the
inertia-buoyancy mode by the SDW mode. Figure 2 shows the temporal
variation of the coupling parameters during the swing of a
perturbation SFH from leading to trailing in the case of
superadiabatic stratification with our chosen $\gamma=0.8$.
Accordingly, from now on we speak only of the convective mode
instead of the more general inertia-buoyancy mode. In this case,
$\hat{\omega}_g^2(t)$ has a negative sign for most of the time (Fig.
1) and determines the shear-modified instantaneous growth rate of
convective instability. It is seen from Fig. 2 that for fixed $K_y$
and $K_z$, these parameters reach their maximal values in the
interval $|t| \lsim 1$, when the time-dependent radial wavenumber is
not large, $|K_x(t)/K_y|=q|t| \lsim 1$, and rapidly decay at $|t|
\gg 1$, when $|K_x(t)/K_y|=q|t| \gg 1$ as well. Thus, because the
coupling parameters $\chi$ evidently also originate from shear, we
can conclude that at given $K_y$ and $K_z$, the influence of shear
on the mode dynamics can be important only for moderate radial
wavenumbers and an efficient energy exchange between the modes and
the mean disc flow should be expected to occur just during this
interval, which we will call a coupling interval. For
$|K_x(t)/K_y|\gg 1$, the coupling parameters are small and hence the
modes become dynamically decoupled from each other and evolve
independently (see the next section). In Fig. 3, the maximal values
(over the interval $|K_x(t)/K_y| \lsim 1$) of the modulus of the
coupling parameters are plotted in the $(K_y,K_z)$-plane. The
maximum values of $|\chi_{sg2}|$, $|\chi_{gs1}|$ and $|\chi_{gs2}|$
are appreciable for of the order of unity or smaller $K_z, K_y$ and
rapidly decay with the increase of these wavenumbers. The maximum of
$|\chi_{sg1}|$ is appreciable over a broader range of wavenumbers,
but it has larger values at smaller $K_y$. Because the coupling
parameters are generally one of the factors determining the
generation of one mode by another, we should expect the efficiency
of coupling of the SDW and convective modes as a function of $K_y$
and $K_z$ to follow, to a certain extent, the same trend as that of
the coupling parameters shown in Fig. 3. Although in a
superadiabatically stratified disc the generation (triggering) of
convective motions by SDWs, determined by the coupling parameters
$\chi_{gs1},\chi_{gs2}$, is in principle possible, here we focus
only on excitation of SDWs by convection, determined by
$\chi_{sg1},\chi_{sg2}$, as it seems more interesting. Besides,
convection can be easily triggered by other mechanisms in discs.

\section{SDW and convective modes -- shear-induced coupling}

In this section, we study the shear-induced dynamics of the
non-axisymmetric SDW and convective modes by numerically solving
equations (32) and (33). At large times/radial wavenumbers
($|K_x(t)/K_y|=q|t|\gg 1$), the adiabatic (WKBJ) condition with
respect to time is satisfied for the mode frequency/growth rate,
\[
|\hat{\omega}_{s,g}'(t)|\ll |\hat{\omega}_{s,g}^2(t)|.
\]
A physical interpretation of this inequality is that the time-scale
of shear-induced variation of the frequency/growth rate is much
larger than the characteristic time-scale (i.e., inverse
frequency/growth rate) of the mode itself. In this adiabatic regime,
$\hat{\omega}_g^2(t)\simeq N_0^2<0$, so it is nearly independent of
time and is much less than the frequency of SDWs, which increases
linearly with time, $\hat{\omega}_s(t)\simeq K(t)\approx qK_y|t|\gg
1$, so that the mode time-scales are well separated (Fig. 1). In
addition, all four coupling parameters $\chi$ are negligible at
these times (Fig. 2). These imply that in the adiabatic regime,
shear plays only a minor role in the dynamics of SDWs and convective
instability. Below we will refer to this asymptotic stage as
adiabatic/non-coupling region in the ${\bf K}$-space. Consequently,
if the convective mode is in the adiabatic region, it will evolve
with time as in the shearless limit, without any exchange of energy
and exciting other modes (i.e., SDWs) in the disc. However, if
$K_x(t)$ crosses the coupling interval $|K_x(t)/K_y| \lsim 1$ during
its drift, where, as we have seen, the influence of shear on the
mode dynamics becomes significant, the excitation of SDWs by
non-axisymmetric convective motions can take place. So, the goal of
this section is to examine this phenomenon in detail.

For further use, we also introduce the mode energies as
\[
E_s\equiv|\Psi_s'|^{2}+\hat{\omega}_s^2(t)|\Psi_s|^2,~~~~E_g\equiv|\Psi_g'|^{2}+\hat{\omega}_g^2(t)|\Psi_g|^2.
\]
for the SDW and convective modes, respectively. In the absence of
shear they are conserved quantities, as it follows from equations
(29). The mode energies are useful in characterising mode dynamics
for asymptotically large times, in the adiabatic regime. In this
regime, the energies vary with time due solely to the mean shear
flow. In particular, the SDW mode energy grows linearly with time,
\[
E_{s}\propto \hat{\omega}_{s}(t)\simeq K(t)\propto |t|,
\]
which follows from equation (32) in the WKBJ regime neglecting the
coupling parameters, which are small in this regime anyway. Such an
asymptotic behaviour of the SDW mode energy will later prove to be a
useful diagnostic in analysing the generation of SDWs. Note also
that at large times $E_s$ is obviously positive definite, whereas
$E_g$ can become negative (Fig. 4), since $\hat{\omega}_g^2\simeq
N_0^2<0$ at these times.

\subsection{Generation of the SDW mode by the convective mode}

We wish to investigate the convective instability in the presence of
Keplerian shear and, specifically, how convective motions can excite
SDWs. To this end, initially at $t=-t_0$, where $t_0\gg 1$ is some
large positive parameter, we impose a tightly leading (i.e., with
$K_x(-t_0)/K_y \ll -1$) SFH of the convective mode on the flow
without any mixture of a SDW mode SFH and follow the subsequent
evolution of the eigenfunctions and perturbed quantities until
$t=t_0$, at which stage their SFHs become tightly trailing (with
$K_x(t_0)/K_y \gg 1$). For numerical integration we use a standard
Runge-Kutta scheme (MATLAB ode45 RK implementation).

To prepare such initial conditions, we make use of the fact that at
$t\ll -1$ the dynamics is adiabatic and the modes do not interact
with each other that, in turn, permits us to pick out (impose) only
the convective mode at the beginning. In this adiabatic/non-coupling
regime, the convective mode is given by the WKBJ asymptotic solution
of the homogeneous part of equation (33), which we take to have the
form
\begin{equation}
\Psi_g=\frac{C_0}{\sqrt{|\hat{\omega}_g(t)}|}{\rm
exp}\left(\int_{-t_0}^t |\hat{\omega}_g(t')|dt'\right) ~~~~ at ~~~~
t\ll -1,
\end{equation}
where $C_0$ is a some arbitrary constant setting the initial value
of $\Psi_g$ at the start of integration at $t=-t_0$. This solution
means that the convective mode is evanescent in the distant past (at
$t\rightarrow -\infty$) and grows with time afterwards. As for the
initial value of the SDW mode eigenfunction, a high-frequency SDW
component is absent at the outset, so that $\Psi_s$ should be given
only by non-oscillatory particular solution, $\Psi_s^{\rm (g)}$, of
equation (32) coming from the source term on the right hand side
that is associated with the convective mode. Asymptotically at $t\ll
-1$, we can ignore the second order time derivative in equation (32)
and represent the initial value of $\Psi_s$ at these times, to a
good approximation, as
\begin{equation}
\Psi_s=\Psi_s^{\rm (g)}\approx
\frac{\chi_{sg1}\Psi'_g+\chi_{sg2}\Psi_g}{\hat{\omega}_s^2(t)}.
\end{equation}
Indeed, as mentioned above, in the adiabatic regime, (${\rm i}$) the
period of the SDW oscillations is much smaller than the
shear/dynamical time, or equivalently $|\hat{\omega}_{s}'| \ll
\hat{\omega}_{s}^2$, ($\rm ii$) the coupling parameters vary a
little during the oscillation period and at the same time (${\rm
iii}$) the frequency of SDWs is much larger than the characteristic
growth rate of the convective mode, $|\hat{\omega}_g|\ll
\hat{\omega}_s$. As a result, the second time derivative of the
particular solution (35) turns out to be much smaller than the right
hand side term of equation (32), thereby validating our
approximation. This slowly varying solution ensures that the
oscillatory SDW mode is absent in the initial conditions. Note also
that as the coupling coefficients are vanishing at $t\rightarrow
-\infty$, we also have $\Psi_s^{\rm (g)}(-\infty)\rightarrow
0,~\Psi_s^{\rm (g)'}(-\infty)\rightarrow 0$. This in fact coincides
with the condition of the absence of SDWs in the shearless limit.
Thus, asymptotic solutions (34) and (35) can be viewed as a
generalisation of the recipe for imposing only the convective mode,
without a mixture of SDWs, on the flow in the absence of shear
(i.e., $\Psi_s(-\infty)=\Psi'_s(-\infty)=0$) to the case of non-zero
shear provided, however, that the adiabatic approximation is still
met, which is always the case at large times.

\begin{figure}
\centering\includegraphics[width=\columnwidth]{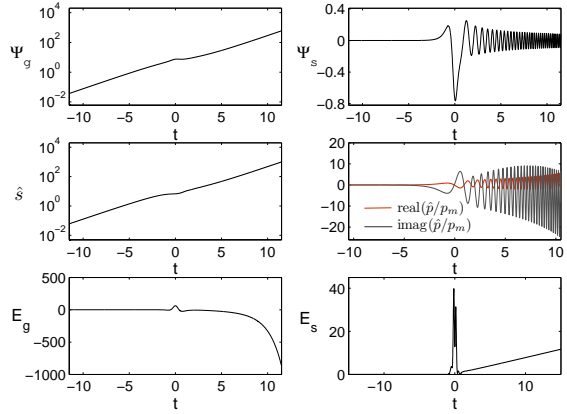}
\caption{Evolution of the eigenfunctions $\Psi_g$ and $\Psi_s$,
entropy $\hat{s}$, normalised perturbed pressure $\hat{p}/p_m$ and
corresponding energies $E_g, E_s$ pertaining to an initially imposed
purely convective mode SFH with $K_y=2, K_z=3$. At the beginning,
$\Psi_g$ and $\hat{s}$ evolve adiabatically, exponentially growing
with the characteristic growth rate $|\hat{\omega}_g|$, while
$\Psi_s$ and $\hat{p}$ are still nearly zero. At around $t=0$, rapid
oscillations abruptly emerge in the evolution of $\Psi_s$ and
$\hat{p}$ that indicate a trailing SFH of the SDW mode being
excited. In addition, the convective mode causes $\hat{p}$ to grow
as well while oscillating. Accordingly, the energy of the SDW mode,
$E_s$, being negligible at $t<0$, after the transient amplification
event in the vicinity of $t=0$, increases linearly with time in the
subsequent adiabatic regime at $t \gg 1$ (which actually starts
already at $t\sim 1$).}
\end{figure}

Figure 4 shows the subsequent temporal evolution corresponding to
initial conditions/solutions (34) and (35). Together with the
eigenfunctions, which more clearly illustrate mode coupling, we also
plot the time-development of the perturbed entropy $\hat{s}$,
normalised pressure $\hat{p}/p_m$ and mode energies $E_g$ and $E_s$.
At the beginning, while the convective mode SFH is still in the
adiabatic region $K_x(t)/K_y=qt\ll -1$ and far from the coupling
region, the eigenfunction $\Psi_g$, following expression (34), grows
exponentially with the instantaneous growth rate $|\hat{\omega}_g|$
and induces a similar behaviour in the entropy, whereas $\Psi_s$,
evolving according to (35), still remains very small and
non-oscillatory. The evolution of the corresponding energy $E_s$ is
similar to that of $\Psi_s$. It is also seen in Fig. 4 that at this
initial adiabatic stage of evolution, the entropy perturbation is
dominant over the pressure perturbation, which, like $\Psi_s$, is
small and aperiodic initially. It can also be shown that the form of
our initial conditions for the convective mode implies $E_g=0$ at
$t\ll -1$, because of the negative $\hat{\omega}_g^2<0$; of course,
this does not mean that the convective mode is zero.

As time passes, the time-dependent radial wavenumber $K_x(t)$ of the
convective mode SFH gradually approaches the coupling region
$|K_x(t)/K_y|\lsim 1$ and begins to cross it. In this region, the
effects of shear come into play: the coupling parameters become
appreciable (Fig. 2) and simultaneously the characteristic
time-scales of the SDW mode, convective mode and the shear time
($=1/q$ in the non-dimensional form) become
comparable,\footnote{This is not the case at large $K_y$ and/or
$K_z$, because the mode time-scales still remain separated even for
$|K_x(t)/K_y|\lsim 1$ and therefore the mode coupling is negligible
at large wavenumbers; see section 3.2.} or equivalently $q \sim
|\hat{\omega}_g| \sim \hat{\omega}_s$ (Fig. 1), also implying that
in the coupling region, the adiabatic condition for the SDW
frequency breaks down, $|\hat{\omega}'_{s}|\sim
|\hat{\omega}_{s}^2|$, that is, the coupling region is also
non-adiabatic. This results in an efficient energy exchange between
these two modes and the background shear flow in this coupling
region. First, at $-1\lsim t<0$, the eigenfunction $\Psi_s$,
pressure $\hat{p}$ and the energies $E_g, E_s$ undergo transient
amplification by extracting energy mainly from the background flow,
but still remain aperiodic. The evolution of $\Psi_s$ has now
deviated from the asymptotic solution (35), because the second
derivative of $\Psi_s$ has become important in equation (32), though
it is still following non-oscillatory $\Psi_s^{\rm (g)}$. Then,
during a short period of time around $t=0$, when $K_x(t)$ crosses
the point $K_x=0$, swinging from negative (leading) to positive
(trailing), rapid oscillations abruptly appear in the evolution of
$\Psi_s$ and the pressure, signaling the generation of a trailing
SFH of the SDW mode. So, now at $t>0$, there are the newly excited
trailing SDW mode SFH and the former convective mode SFH. It can be
said that the convective mode, in some sense, acts as a mediator
between SDWs and the disc shear flow. The energy needed for the SDW
excitation is mainly extracted from the shear with the help of the
convective mode. We would also like to note that, as mentioned
above, because the time-scales of the convective and excited SDW
modes are comparable, strictly speaking, these modes cannot be well
distinguished from each other in the coupling region. In other
words, we have a mixture -- `convective-SDW' mode -- at intermediate
times, $|t| \lsim 1$. Then, on leaving the coupling region, $K_x(t)$
moves into the next adiabatic region $K_x(t)/K_y\gg 1$ where the
linear dynamics of SDWs and convection become decoupled with no
further energy exchange, so the modes thus acquire truly separate
identities. The mode time-scales has now been separated (Fig. 1):
the pressure $\hat{p}$ rapidly oscillates as a result of the new SDW
component in $\Psi_s$ with the frequency $\hat{\omega}_s(t)$, which
increases linearly with time and soon becomes much larger than the
growth rate, $|\hat{\omega}_g|$, of the convective instability. The
eigenfunction $\Psi_g$ continues to grow exponentially with this
growth rate and causes a similar growth in the entropy. As typical
of SDWs in the adiabatic regime, the corresponding energy $E_s$
increases linearly with time solely due to the background flow.

\begin{figure*}
\includegraphics[width=\textwidth, height=0.65\textwidth]{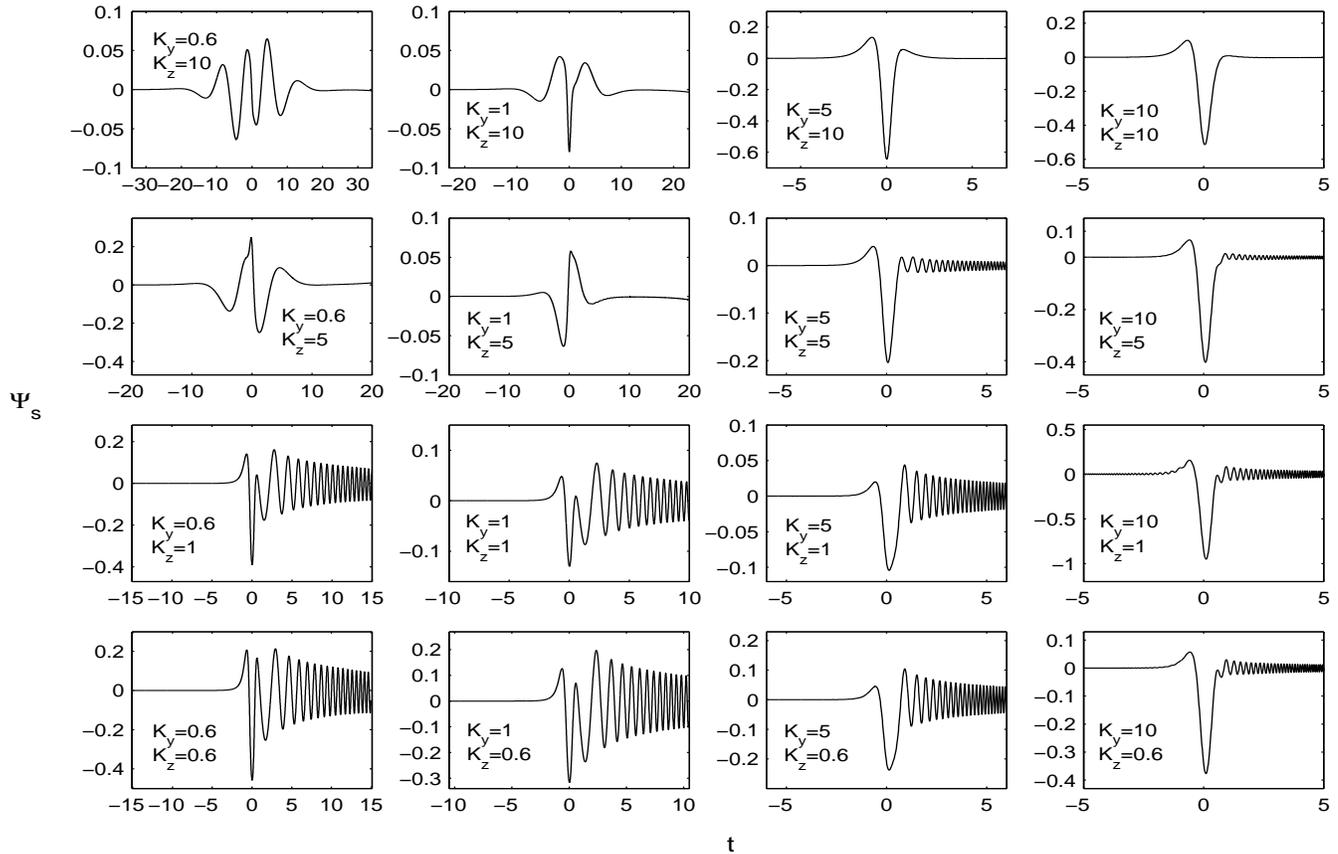}
\caption{Evolution of $\Psi_s$ corresponding to an initially imposed
convective mode SFHs with various $K_y$ and $K_z$. Appreciable
oscillations in the time-development of $\Psi_s$ appear and,
therefore an efficient SDW mode excitation occurs, at $K_y,K_z \lsim
1$. By contrast, at large $K_y$ and/or $K_z$, $\Psi_s$ undergoes
only small transient variation (growth) in the interval $|t|\lsim 1$
with very weak, almost no SDW generation.}
\end{figure*}

Finally, we would like to stress again that the studied here linear
coupling of the SDW and convective modes caused by shear is
essentially of the same nature as other linear mode coupling
phenomena occurring in disc flows with Keplerian shear: coupling of
vortical and wave modes, that is, the generation of SDWs by vortices
\citep[][HP09a]{Bodo_etal05,Johnson_Gammie05a,Mamatsashvili_Chagelishvili07},
coupling of vortices and inertia-gravity waves
\citep{Tevzadze_etal03,Tevzadze_etal08} and coupling of baroclinic
and SDW modes \citep{Tevzadze_etal10}. Thus, the linear mode
coupling is a generic phenomenon inevitably taking place whenever
the mean flow is inhomogeneous (i.e., the velocity profile has a
non-zero shear) and all these cases are its special manifestations
for particular physical settings. In this connection, we should
mention that it also occurs in MHD shear flows, where in general
there are larger number of mode branches with different time-scales
and coupling among them becomes quite complex
\citep{Chagelishvili_etal96}. For example,
\citet{Heinemann_Papaloizou09b} described the generation of SDWs by
the vortical mode in a MRI-turbulent disc shear flow. The coupling
among different MHD modes can also be seen in the linear analysis of
non-axisymmetric MRI by \citet{Balbus_Hawley92}, however, the
authors do not identify it as such. Thus, the primary effect of
shear of the disc's differential rotation on the perturbations
dynamics is that it couples, or introduces a new channel of energy
exchange among different types of perturbation modes existing in the
disc as well as with the disc flow. In the present case, the linear
dynamics of the convective mode is accompanied by the generation of
high-frequency SDWs during a finite-time interval as SFHs swing from
leading to trailing. As noted before, the shearing wave (non-modal)
approach, because of not involving spectral expansion in time,
allows us to trace an entire temporal evolution of perturbations and
thus to reveal new aspects of non-axisymmetric convection in disc
shear flows. This, in turn, can provide further insight into the
role of convection and its coupling with SDWs in angular momentum
transport, especially now when the capability of convection to
transport angular momentum outwards in discs has recently been
established via numerical simulations (we address the transport
properties of the modes in section 4). Previous linear studies of
non-axisymmetric convection have been either in the framework of the
modal approach \citep[][]{Lin_etal93} or using the shearing wave
(non-modal) approach
\citep[RG92;][]{Korycansky92,Brandenburg_Dintrans06} as here, but
without identification and characterisation of the mode coupling
process. As pointed out in the Introduction, the results of the
modal approach are actually applicable for asymptotically large
times and therefore tend to overlook dynamical effects at
intermediate times arising as a result of the flow
non-normality/shear. Since the excitation of SDWs by convection is
just one of such examples occurring during a limited time interval,
in fact it cannot be captured in the framework of the modal
approach.

\subsection{Generation of SDWs for various $K_y$ and $K_z$}

Here we examine how the efficiency of the above-described SDW
generation process depends on the azimuthal, $K_y$, and vertical,
$K_z$, wavenumbers. Figure 5 shows the evolution of $\Psi_s$ under
the same initial conditions (34) and (35) as above, that is,
consisting of an initially imposed only tightly leading convective
mode SFH at various $K_y$ and $K_z$, without a mix of the SDW mode
SFH. For large $K_y$ and/or $K_z$ (i.e., when the azimuthal and/or
vertical wavelengths of the modes are much smaller than the disc
scale height), starting out with very small values, $\Psi_s$ mainly
undergoes only transient variation in the coupling region (at $|t|
\lsim 1$) followed by very weak, almost no generation of
high-frequency oscillations, i.e., SDW component. The reason for
such a behaviour is the following. At large $K_y$ and/or $K_z$, the
time-scales of the SDW and convective modes remain well separated
during an entire course of evolution, not only at large times, and
the adiabatic condition for the SDW frequency holds. As a result,
non-oscillatory particular solution (35), valid for the initial
adiabatic stage of evolution at $t\ll -1$, in fact continues to hold
at all times to a good approximation (because its second time
derivative always remains negligible in equation 32) and,
consequently, an efficient coupling between the SDW and convective
modes is not feasible at large $K_y$ and/or $K_z$. So, in Fig. 5,
the time-development of $\Psi_s$ at these wavenumbers is given by
this particular solution which does not involve any oscillations.
(To be more precise, the SDW mode is still being generated, but with
a very small, as evident from the panels with
$K_y=5,K_z=5;~K_y=10,K_z=5;~K_y=10,K_z=1;~K_y=10,K_z=0.6$ of Fig. 5
and also from Fig. 6.) Thus, the regime of large azimuthal and/or
vertical wavenumbers is a weak coupling regime, where shear plays
only a minor role in the mode dynamics. By contrast, for smaller
$K_y,K_z \lsim 1$ (i.e., for wavelengths comparable to the disc
scale height) the favourable conditions for mode coupling --
violation of the adiabatic condition and comparable time-scales of
the SDW and convective modes and the shear time -- can occur in the
vicinity of $t=0$, leading to an efficient generation of the SDW
mode by the convective one, as described in section 3.1.

We can quantitatively characterise the mode coupling as follows. At
$t>0$, the total solution for $\Psi_s$ consists of the two
components
\[
\Psi_s=\Psi_s^{\rm (w)}+\Psi_s^{\rm (g)},
\]
where $\Psi_s^{\rm (g)}$, as before, is the non-oscillatory, slowly
varying particular solution of equation (32) due to the convective
mode and $\Psi_s^{\rm (w)}$ is an oscillatory component related to
the SDW mode generated by this particular solution. So, the relative
intensity of the wave generation process can be quantified by
comparing the value of $\Psi_s^{\rm (w)}$ to that of $\Psi_s^{\rm
(g)}$. To do this, we introduce the parameter
\[
\epsilon\equiv\frac{\max_{t>0}|\Psi_s^{\rm
(w)}(t)|}{\max_{t>0}|\Psi_s^{\rm (g)}(t)|},
\]
which is the ratio of the maximal values of $|\Psi_s^{\rm (w)}|$ and
$|\Psi_s^{\rm (g)}|$ over time (typically, both these functions
achieve their maximal values in the vicinity of $t=0$, as seen in
Fig. 5). Obviously, with our initial conditions (34) and (35), this
ratio does not depend on the arbitrary parameter $C_0$ and is a
function of $K_y$ and $K_z$ only, which is plotted in Fig. 6. It is
seen from this figure that at large $K_y$ and/or $K_z$, as expected,
$\epsilon \lsim 0.1$ is small, implying that the non-oscillatory
particular solution dominates over the oscillatory SDW mode
solution, i.e., wave generation is weak. For $K_y,K_z\lsim 1$,
$\epsilon \sim 1$ implying that the wave component is comparable to
the non-oscillatory solution and therefore the SDW generation is
appreciable. This also confirms the situation in Fig. 5.

\begin{figure}
\centering\includegraphics[width=0.95\columnwidth]{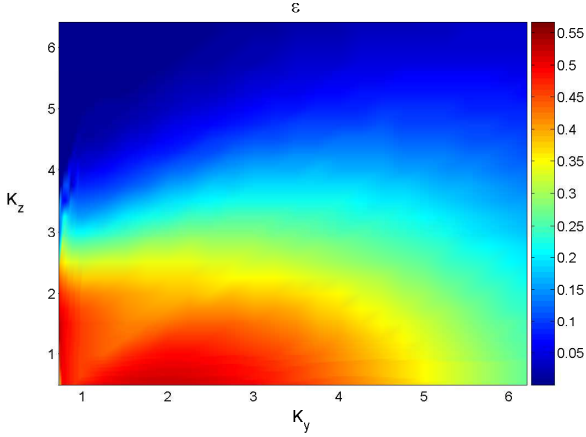}
\caption{Ratio, $\epsilon$, of the maximum values of the SDW mode
solution to its generator non-oscillatory particular solution of
equation (32) due to the convective mode, plotted as a function of
$K_y$ and $K_z$.}
\end{figure}

\subsection{Amplitudes of generated SDWs}

We can go further and calculate the amplitudes of SDWs generated by
the convective mode, which also represent a measure of the coupling
efficiency between these two modes. In this section, we choose the
constant $C_0$ in the WKBJ solution (34), describing an initially
tightly leading convective mode SFH at $t \ll -1$, such that to have
\[
\Psi_g=\frac{C_0}{\sqrt{|\hat{\omega}_g(t)|}}{\rm
exp}\left(\int_{-t_0}^t |\hat{\omega}_g(t')|dt'\right)=
\frac{1}{\sqrt{|N_0|}}{\rm e}^{|N_0|t}.
\]
An advantage of this form is that the calculated below SDW
amplitudes depend only on $K_y$ and $K_z$ and are independent of the
choice of the starting point $-t_0$ (as long as $t_0\gg 1$). In this
initial adiabatic regime, as before, $\Psi_s=\Psi_s^{\rm (g)}$,
because SDWs are absent at the outset. Then, in the non-adiabatic
region at $|t| \lsim 1$, the mode coupling is at work and
subsequently, at around $t=0$, the SDW mode SFH abruptly emerges.
After crossing the non-adiabatic/coupling region, in the next
adiabatic region at $t\gg 1$, there are the former convective mode
SFH and the generated by it SDW mode SFH, both with tightly trailing
orientation. So, now, as noted above, the full solution for $\Psi_s$
is the sum of the following parts:
\begin{multline*}
\Psi_s=\Psi_s^{\rm (w)}+\Psi_s^{\rm
(g)}=\\=\frac{A}{\sqrt{\hat{\omega}_s(t)}}{\rm exp}\left(-{\rm
i}\int^t_0\hat{\omega}_s(t')dt'\right)+\\+\frac{A^{\ast}}{\sqrt{\hat{\omega}_s(t)}}{\rm
exp}\left({\rm i}\int^t_0\hat{\omega}_s(t')dt'\right)+\\
+\frac{\chi_{sg1}\Psi'_g+\chi_{sg2}\Psi_g}{\hat{\omega}_s^2(t)}~~~at~~~t\gg
1,
\end{multline*}
where the first two terms are oscillatory WKBJ solutions of the
homogeneous part of equation (32) that correspond to the excited
SDWs. Since the initial conditions are real, these solutions come in
complex conjugate pairs with different signs of frequency and with
amplitudes $A$ and its complex conjugate $A^{\ast}$. Thus, the SDW
mode SFH generated by the convective mode is actually a
superposition of two SFHs corresponding to SDWs propagating in
opposite directions. In other words, convection always pairwise
excites SDWs that propagate oppositely, similar to SDWs generated by
the vortical mode in 2D discs (see e.g., HP09a). The third term is
again the non-oscillatory slowly varying particular solution due to
the convective mode.

\begin{figure}
\centering\includegraphics[width=\columnwidth]{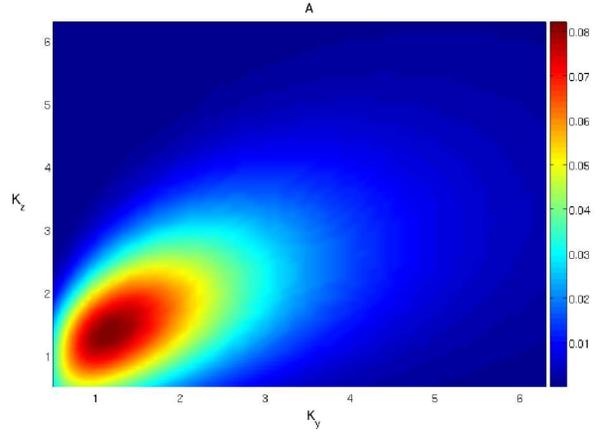}
\caption[Absolute value of the amplitude of SDWs generated by the
convective mode, $|A|$, as a function of $K_y$ and $K_z$]{Absolute
value of the amplitude, $|A|$, of SDWs generated by the convective
mode as a function of $K_y$ and $K_z$. It achieves the largest value
$|A|_{max}=0.082$ at $K_{y,m}=1.2,K_{z,m}=1.3$ and decreases at
smaller and larger $K_y,K_z$, implying also the decrease in the mode
coupling efficiency at these wavenumbers.}
\end{figure}

Figure 7 shows the dependence of the absolute value of the
amplitude, $|A|$, on $K_y$ and $K_z$. When either of these
wavenumbers is large, $|A|$ is small, because, as shown above, the
SDW generation/mode coupling is inefficient. With decreasing $K_y$
and $K_z$, as expected, the amplitude increases and attains its
maximum values at $K_y,K_z \lsim 1$, because an appreciable SDW
generation takes place at such wavenumbers, as seen in Figs. 5 and
6. Note that $K_y,K_z \lsim 1$ is the region where the coupling
parameters $\chi_{sg1},\chi_{sg2}$ are appreciable too (see Fig. 3).
Comparing Figs. 6 and 7, we also see that the amplitude $|A|$ and
$\epsilon$, which is a measure of the mode coupling efficiency,
behave with $K_y$ and $K_z$ more or less similarly.

\section{Angular momentum transport by the convective and SDW modes}

In this section, we analyse the angular momentum transport
properties of the SDW and convective modes. Specifically, we are
interested in the radial component of angular momentum flux carried
by these two modes, because it determines the radial transport of
mass and hence accretion onto the central star. The angular momentum
conservation law for linear perturbations in the shearing box model
follows from the invariance of the system under translations in the
azimuthal $y$-direction. It can be derived by application of
Noether's theorem to the second order Lagrangian density \citep[RG2;
HP09a;][]{Narayan_etal87}
\begin{multline*}
L=\frac{\rho_0}{2}\left(\frac{D {\boldsymbol
\xi}}{Dt}\right)^2-\frac{p_0}{2}\left[(\gamma-1)(\nabla\cdot
{\boldsymbol \xi})^2+\frac{\partial \xi_i}{\partial
x_j}\frac{\partial \xi_j}{\partial x_i}\right]+\\+\rho_0({\bf
\Omega}\times {\boldsymbol \xi})\cdot \frac{D {\boldsymbol
\xi}}{Dt}+q\rho_0 \Omega^2\xi_x^2,
\end{multline*}
in this direction and yields
\[
\frac{D}{Dt}\left(\frac{\partial L}{\partial
(D\xi_i/Dt)}\frac{\partial \xi_i}{\partial y}\right)+\frac{\partial
}{\partial x_j}\left(\frac{\partial L}{\partial (\partial
\xi_i/\partial x_j)}\frac{\partial \xi_i}{\partial
y}-L\delta_{j2}\right)=0,
\]
where $\boldsymbol \xi$ is the displacement vector related to the
perturbed velocity by
\[
u_x=\frac{D \xi_x}{Dt}, ~~~ u_y=\frac{D \xi_y}{Dt}+q\Omega\xi_x,
~~~u_z=\frac{D \xi_z}{Dt},
\]
and $\delta_{ij}$ is the Kronecker delta and the summation is
assumed over the repeated indices $i,j=1,2,3$ corresponding to
$(x_1,x_2,x_3)\equiv (x,y,z)$. (It can also be readily shown that
this Lagrangian density through variation gives the original set of
linear perturbation equations 9-13). Since we are mainly concerned
with the radial $x$-component of the angular momentum flux, we can
average this equation over both $y$- and $z$-coordinates. After
averaging, the canonical angular momentum density in brackets in the
first term, as demonstrated by \cite{Narayan_etal87}, actually
coincides with the density of the radial component of true physical
angular momentum, but taken with the minus sign and without fiducial
radius $r_0$ that does not play any role in the local analysis. So,
we write for the radial component, $B$, of angular momentum density
\[
B=-\left<\frac{\partial L}{\partial (D\xi_i/Dt)}\frac{\partial
\xi_i}{\partial y}\right>_{yz},
\]
where the angle brackets denote averaging over $y$ and $z$. On
similarly averaging the second term and taking it with the minus
sign, we obtain the desired radial component of the angular momentum
flux
\[
F_x=-\left<\frac{\partial L}{\partial (\partial \xi_i/\partial
x)}\frac{\partial \xi_i}{\partial
y}\right>_{yz}=-\left<p'\frac{\partial \xi_x}{\partial
y}\right>_{yz},
\]
where the pressure perturbation $p'$ (as used in equations 9-13, the
prime is not time-derivative in this case) is related to the
displacement vector through $p'=-\rho_0c_s^2\nabla\cdot{\boldsymbol
\xi}+\rho_0g\xi_z$. After that the angular momentum conservation
takes more compact form
\[
\frac{\partial B}{\partial t}+\frac{\partial F_x}{\partial x}=0.
\]
For the angular momentum flux associated with an individual SFH
given by (15) and (16) after spatial averaging we get
\[
F_x=\frac{{\rm
i}}{4}K_y[\hat{p}(t)\hat{\xi}_x^{\ast}(t)-\hat{\xi}_x(t)
\hat{p}^{\ast}(t)],
\]
where again $\hat{p}(t)$ and $\hat{\xi}_x(t)$ are the amplitudes of
SFHs depending only on time with $\hat{p}^{\ast}(t)$ and
$\hat{\xi}_x^{\ast}(t)$ being their respective complex conjugates.
The velocity and displacement amplitudes are related by
$\hat{u}_x(t)=d\hat{\xi}_x(t)/dt$. We normalise displacement
${\boldsymbol \xi}$ as other scale-lengths above,
$\Omega{\boldsymbol \xi}/c_s \rightarrow {\boldsymbol \xi}$, and
pressure and flux as $\hat{p}/p_m\rightarrow \hat{p}, F_x/p_m
\rightarrow F_x$. Since, we have two modes -- SDWs and convection,
we decompose $\hat{p}$ and $\hat{\xi}_x$ into two parts
corresponding to these modes
\[
\hat{p}=\hat{p}_s+\hat{p}_g, ~~~
\hat{\xi}_x=\hat{\xi}_{x,s}+\hat{\xi}_{x,g},
\]
where $\hat{p}_s$ and $\hat{\xi}_{x,s}$ are related to the SDW
component and, therefore, are rapidly oscillating in time with the
frequency of the latter, while $\hat{p}_g$ and $\hat{\xi}_{x,g}$ are
related to the convective mode and vary on its corresponding
time-scale, relatively slowly compared with the SDW mode. So, the
total flux
\begin{multline*}
F_x=\frac{\rm
i}{4}K_y[(\hat{p}_s+\hat{p}_g)(\hat{\xi}_{x,s}^{\ast}+\hat{\xi}_{x,g}^{\ast})-\\
-(\hat{\xi}_{x,s}+\hat{\xi}_{x,g})(\hat{p}_s^{\ast}+\hat{p}_g^{\ast})]
\end{multline*}
is also of oscillatory type because of the presence of the SDW mode,
but we can conveniently smooth it by averaging over the oscillation
period much smaller than the growth time of the convective mode (we
denote the time-averaging with angle brackets without subscript). As
a result, we have
\begin{multline}
F_x=\frac{\rm
i}{4}K_y\langle\hat{p}_s\hat{\xi}_{x,s}^{\ast}-\hat{\xi}_{x,s}
\hat{p}_s^{\ast}\rangle+\frac{\rm
i}{4}K_y\langle\hat{p}_g\hat{\xi}_{x,g}^{\ast}-\hat{\xi}_{x,g}
\hat{p}_g^{\ast}\rangle.
\end{multline}
Thus, all the cross-products vanish after time-averaging and the
total angular momentum flux appears as a sum of angular momentum
fluxes
\[
F_{x,s}=\frac{\rm
i}{4}K_y\left<\hat{p}_s\hat{\xi}_{x,s}^{\ast}-\hat{\xi}_{x,s}
\hat{p}_s^{\ast}\right>,
\]
\[
F_{x,g}=\frac{\rm
i}{4}K_y(\hat{p}_g\hat{\xi}_{x,g}^{\ast}-\hat{\xi}_{x,g}
\hat{p}_g^{\ast})
\]
related, respectively, to the SDW and convective modes (we have
omitted the angle brackets in $F_{x,g}$, as it does not change much
during the oscillation period). In an analogous calculation of the
angular momentum transport by non-axisymmetric shearing waves, RG92
obtained a rapidly oscillating, on average negative, angular
momentum flux and attributed it only to the convective mode without
separating/analysing the contribution from SDWs. In fact, this
contribution is present, because, as demonstrated here, SDWs are
inevitably generated by the convective mode due to background shear.
So, decomposition (36) allows us to analyse the angular momentum
fluxes carried by each mode individually and contrast them.

\begin{figure}
\centering\includegraphics[width=\columnwidth]{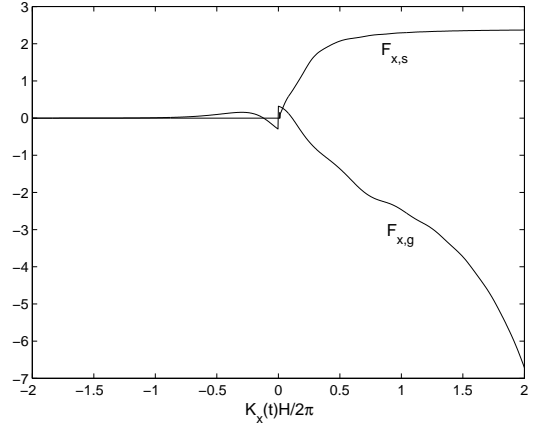}
\caption{Radial angular momentum fluxes associated with the
convective mode, $F_{x,g}$, and with the generated by it the SDW
mode, $F_{x,s}$, for $K_{y,m}=1.2,K_{z,m}=1.3$. The jump of
$F_{x,s}$ and $F_{x,g}$ in the immediate vicinity of $t=0$ is due to
the abrupt emergence of the SDW mode.}
\end{figure}

In Fig. 8, we show the time-history of the fluxes $F_{x,s}$ and
$F_{x,g}$ when a tightly leading purely convective mode SFH is
imposed initially. At $t<0$, before SDW mode generation, angular
momentum transport is solely due to the convective mode SFH and is
still small and positive. Then, in the vicinity of $t=0$, trailing
SDW mode SFH emerges, giving rise to its associated flux $F_{x,s}$
at $t>0$, which initially increases and soon settles to a positive
and constant value. Thus, trailing SDWs always transport angular
momentum outwards. (A similar result was also obtained by HP09a for
SDWs generated by the vortical mode.) As for the angular momentum
flux of the convective mode, during a short period of time after
$t=0$, it is still positive, but quickly switches to increasingly
negative due to the exponential growth of $\Psi_g$, and dominates
over the positive flux of the SDW mode with time. Because of the
relatively short time-scale of the wave generation process in the
immediate vicinity of $t=0$, the variations of $F_{x,s}$ and
$F_{x,g}$ also appear sharp around $t=0$. However, as demonstrated
by \citet{Lesur_Ogilvie10} and \citet{Kapyla_etal10}, this flux of
angular momentum associated with the convective mode, being negative
in the linear regime, can actually change sign (direction) in the
non-linear regime. We also see from Fig. 8 that shortly after the
SDW mode generation, at $K_x(t)H/2\pi\lsim 1$ (i.e., when the radial
wavelength is still larger than or equal to the scale height), its
corresponding flux, $F_{x,s}$, achieves a maximum value comparable
to, though still less by absolute value, than that of the convective
mode, $F_{x,g}$. As \citet{Heinemann_Papaloizou09b} showed, just
these values of the angular momentum flux of SDWs during the time
when their time-dependent $K_x(t)$ are still in the range
$K_x(t)H/2\pi\lsim 1$, are important and determine transport in the
non-linear regime, because SDWs with radial wavenumbers larger than
this are subsequently damped via shock formation in the trailing
phase.\footnote{But overall, wave perturbations and associated
transport are not damped in a fully developed turbulence, as
successive leading SFHs are regenerated due to non-linearity,
maintaining a steady angular momentum flux.} Thus, SDWs generated by
convection are, in principle, capable of fully contributing to the
transport before undergoing significant damping. Based on this, in
the non-linear regime, we may expect that the positive flux due to
SDWs will aid and enhance the outward angular momentum transport due
to the convective mode alone, although a further non-linear study is
required to ascertain this.

\section{Summary and discussion}

In this paper, we have performed a detailed investigation of a new
linear mechanism of spiral density wave excitation by vertical
thermal convection in compressible Keplerian discs with
superadiabatic vertical stratification, using the local shearing box
approach. The wave excitation results from the velocity shear of the
disc's Keplerian differential rotation. As is usually done in the
shearing box, perturbations were decomposed into spatial Fourier
harmonics, or shearing plane waves. The temporal evolution of the
amplitudes of these waves was followed by numerical integration of
the linearised hydrodynamical equations of the shearing box. Only
non-axisymmetric perturbations were considered, as only for them the
effects of shear are important. Three basic types of perturbation
modes can be distinguished in the considered system: SDWs due mainly
to compressibility, the convective mode due mainly to the negative
vertical entropy gradient, and the vortical mode due to the combined
action of vertical stratification and the Coriolis force. However,
in the present case of non-self-gravitating discs with
superadiabatic vertical structure, the main energy-carrying mode is
convection. In such a setup, the vortical mode can be neglected and
hence we have concentrated primarily on the dynamics of SDWs and the
convective mode, which both have zero potential vorticity. We first
characterised the properties of the SDW and convective modes by
deriving the dispersion relations in the presence of disc rotation
and stratification, but neglecting shear of disc flow. In this limit
of rigid rotation, the modes evolve independently, that is, once
either of these modes has been initially excited, it does not lead
to the excitation of another. We then demonstrated that on taking
into account differential rotation/shear, an initially tightly
leading SFH of the convective mode evolving in the disc flow, on
becoming trailing, generates a corresponding SFH of SDWs. The main
mechanism of wave generation is the following. Because of shear, the
radial wavenumber of SFH varies with time, making the characteristic
time-scales of these modes time-dependent as well. As a result,
during a short period of time when SFHs swing from leading to
trailing, the mode time-scales can become comparable to each other
and to the shear time, thereby making an efficient energy exchange
between the SDW mode, convective mode and the mean disc flow
possible. Consequently, SDWs are generated in this process, mainly
at the expense of shear flow energy, with the help of the convective
mode. We also quantified the efficiency of wave generation at
different azimuthal and vertical wavenumbers and found that it is
maximal when these length-scales are comparable to the disc scale
height. This, in turn, implies that SDWs excited by convection have
weak vertical dependence, similar to SDWs generated by vortices (see
HP09a). We calculated the angular momentum flux associated with
non-axisymmetric SDWs generated by the convective mode and found
that it is positive, i.e., transport is outward as opposed to that
of the convective mode.

Studied here linear coupling of SDWs and convection is basically
similar in nature to the linear coupling of SDWs and vortices that
has already been extensively studied
\citep[][HP09a]{Bodo_etal05,Mamatsashvili_Chagelishvili07} and
represents a special manifestation of a more general phenomenon of
shear-induced linear coupling of perturbation modes inevitably
taking place in any flow with an inhomogeneous velocity profile, or
shear. Thus, although convection is not generally shear-driven -- in
that it does not directly tap energy from shear but from the
unstable superadiabatic thermal distribution -- its dynamics is
still affected by shear and shear-induced coupling to SDWs, in some
sense, makes convection a participant in kinematic processes as
well.

Shearing box simulation of compressible MRI-turbulence in a
Keplerian disc indicates that the angular momentum transport due to
non-axisymmetric SDWs generated by vortical perturbations
constitutes a significant fraction of that associated with the total
turbulent Reynolds stresses \citep{Heinemann_Papaloizou09b}. In
other words, a purely hydrodynamic part of transport is mostly due
to these SDWs and is quite appreciable. As mentioned in the
Introduction, in their non-linear simulations,
\citet{Lesur_Ogilvie10} demonstrated that, contrary to previous
results, in fact vertical convection in protoplanetary discs has a
non-axisymmetric structure and is thus able to transport angular
momentum outwards. But the magnitude of the corresponding $\alpha$
reported in their simulations is still small ($\lsim 10^{-4}$).
However, these simulations were performed in the
Boussinesq/incompressible limit, where high-frequency SDWs are
filtered out. In a related study, \cite{Kapyla_etal10} solved full
hydrodynamical equations in the shearing box with superadiabatic
stratification without explicitly making the assumption of
incompressibility, but the parameter regime considered was such as
to give small Mach numbers. So, the effects of compressibility and,
hence of SDWs, were probably small in their simulations and were not
particularly addressed. As noted above, SDWs are generally known to
be able to enhance angular momentum transport rates. For example,
\citet{Johnson_Gammie05b} demonstrated that compressible simulations
of vortices in 2D discs yield at least an order of magnitude larger
transport rate, which is primarily due to shocks of SDWs emitted by
vortices, compared with that in the incompressible case, which is
due to vortices only. Based on this property, we might anticipate a
similar situation if compressibility is taken into account in
simulations of convectively unstable discs, where SDWs generated by
convective motions in the non-linear regime could also boost outward
angular momentum transport due to convection alone, although this
needs to be further investigated in greater detail in future
numerical studies. We emphasise once more that the described here
shear-induced coupling between the SDWs and convection occurs only
when these modes are non-axisymmetric. So, in the earlier
simulations of \citet{Cabot96} and \citet{Stone_Balbus96}, where
vertical convection had an axisymmetric structure, SDW generation
could not be observed in spite of moderate Mach numbers associated
with convective motions. Another point, which also merits
investigation is where in the disc SDWs generated by convection
primarily cause dissipation through shock formation \citep[see
also][]{Stone_Balbus96}, because in all the above-mentioned
simulations convection was not generated self-consistently, but
maintained by an artificially imposed heat flux. This becomes even
more interesting in the light of the result obtained here that in
the linear regime SDWs do not vary much with height. If the shocks
of SDWs dissipate near the surface, this may stifle convection via
reducing the negative entropy gradient, while if -- near the
midplane, this can then sustain convection.

Finally, we would like to mention that strong vertical convection
also occurs in FU Orionis systems during outbursts as a result of
hydrogen ionisation and is one of the important factors for
understanding the nature of the outburst phenomenon
\citep{Zhu_etal09}. In this case, as shown by Zhu et al., convection
is confined to the inner disc, extends over the whole vertical
height and causes strong density fluctuations, because the Mach
number of the convective eddies/motions is of the order of unity. In
other words, convection is compressible, so the question of the SDW
excitation and the role of wave transport in the outburst process
seems relevant for further investigation.

%%%%%%%%%%%%%%%%%%%%%%%%%%%%%%%%%%%%%%%%%%%%%%%%%%%%%%%%%%%%%%%%%%%%%%%%%%%%%%%%%%%%%%%%%%%%%%

\section*{Acknowledgments}
G.R.M. would like to acknowledge the financial support from the
Scottish Universities Physics Alliance (SUPA). He thanks G. D.
Chagelishvili and A. G. Tevzadze for critically reading the
manuscript. The useful and constructive comments from the anonymous
referee, that improved the presentation of our work, are also much
appreciated.

\bibliographystyle{mn2e}
\bibliography{biblio}
\end{document}